\documentclass[11pt]{article}

\usepackage{amssymb}
\usepackage{epsfig}
\usepackage{graphics}
\usepackage{graphicx}
\usepackage{color}

\usepackage[latin1]{inputenc}
\usepackage[english]{babel}

\usepackage{amssymb}
\usepackage{amsmath}
\usepackage{amsfonts}

\numberwithin{equation}{section}

\input amssym.def
\input amssym.tex

\usepackage{booktabs}
\usepackage{ctable}
\usepackage{subfigure}
\usepackage{float}
\usepackage{layout}
\usepackage{fancyhdr}
\usepackage{color}
\usepackage{eurosym}
\usepackage{wrapfig}  

\usepackage{xr}

\def\I{\mathrm{1\!l}}
\def\P{{\mathbb P}}
\def\R{{\mathbb R}}
\def\E{{\mathbb E}}

\def\N{{\mathbb N}}
\def\cl#1{{\cal{ #1}}}

\def\dT{\Delta T}
\def\cvd{$\quad\Box$\medskip}

\def\<{\langle}
\def\>{\rangle}

\def\coeffbin#1#2{\Big(\!\begin{array}{c}#1\cr #2\end{array}\!\Big)}

\def\Z{{\mathbb{Z}}}

\def\e{\mathrm{\bf e}}
\def\v{\mathrm{\bf v}}
\def\la{\ell^2_a}
\def\M{\mathcal{M}}

\newtheorem{theorem}{Theorem}[section]

\newtheorem{lemma}[theorem]{Lemma}

\newtheorem{proposition}[theorem]{Proposition}
\newtheorem{remark}[theorem]{Remark}

\newcommand{\Dy}{\Delta y}
\newcommand{\Vunita}{{\bf{1}}}

\def \Frac{\displaystyle\frac}

\def\proof{\textbf{Proof.} }

\date{}

\begin{document}

\parindent 0pt

\author{{\sc Maya Briani}\thanks{%
Istituto per le Applicazioni del Calcolo, CNR Roma - {\tt m.briani@iac.cnr.it}}\\
{\sc Lucia Caramellino}\thanks{%
Dipartimento di Matematica,
Universit\`a di Roma Tor Vergata - {\tt caramell@mat.uniroma2.it}; corresponding author.}\\
{\sc Antonino Zanette}\thanks{%
Dipartimento di Scienze
Economiche e Statistiche,
Universit\`a di Udine - {\tt antonino.zanette@uniud.it}}}

\title{
\bf
A hybrid tree-finite difference approach\\
for the Heston model
}

%
\maketitle
\begin{abstract}\noindent{\parindent0pt
We propose an efficient hybrid tree-finite difference method in order to approximate the Heston model. We prove the convergence by embedding the procedure in a bivariate Markov chain and we study the convergence of European and American option prices. We finally provide numerical experiments that give accurate option prices in the Heston model, showing the reliability and the efficiency of the algorithm.}
\end{abstract}




\section{Introduction}
The Black-Scholes model was the most popular model for derivative pricing and hedging, although it has shown several problems with capturing dramatic moves in financial markets.
In fact, the assumption of a constant volatility in the Black-Scholes model
over the lifetime of the derivative is not realistic.  As an
alternative to the Black-Scholes model, stochastic volatility models
emerged. The Heston model \cite{hes} is perhaps the most popular stochastic volatility model, allowing one to obtain closed-formulae in the European case using Fourier transform. In the American option pricing case, the main algorithms turn out to be tree methods, Fourier-cosine methods and finite difference methods. Approximating trees for the Heston model have been considered in different papers, see e.g. \cite{LeisenTreeHeston},   \cite{FlorescuViens1} ,
 \cite{FlorescuViens2},  \cite{HilliardSchwartz},
 \cite{GuanXiaoqiang}.
The tree approach of Vellekoop and Nieuwenhuis \cite{vn} actually  provides
at our knowledge the best tree procedure in the literature.
They use an approach which is based on a
modification of an explicitly defined stock price tree where the
number of nodes grows quadratically in the number of time steps.
Fang and Oosterlee \cite{fo} use a Fourier-cosine series expansion
approach for pricing Bermudan options under the Heston model.
As for finite difference methods for solving the
parabolic partial differential equation associated to the option pricing
problems, they can be based on implicit, explicit or alternating direction
implicit schemes. The implicit scheme requires to solve a sparse
system at each time step.
Clarke and Parrott \cite{ClarkeParrott} and Oosterlee \cite{Oosterlee}  formulate the American put
pricing problem as a linear complementarity problem (LCP)  and use an
implicit finite difference scheme combined with a multigrid procedure, whereas
Forsyth, Vetzal and Zvan \cite{ForsythVetzalZvan} use a penalty method.
The explicit scheme is a quick approach although it requires small time
steps to retain the stability. This leads to a large number of time
steps and is not economic in computation. The ADI schemes are good alternative methods.
For example,  Hout and Foulon \cite{hf} investigate four
splitting schemes of the ADI type for solving the PDE Heston equation:
the Douglas scheme, the Craigh-Sneyd scheme, the Modified Craigh-Sneyd
scheme and the Hundsdorfer-Verwer scheme.
Ikonen and Tovainen \cite{it} propose a componentwise splitting method for
pricing American options in the Heston model. The linear
complementarity problem associated to the American option problem is
decomposed into a sequence of five  one-dimensional LCP's problems at
each time step. The advantage is that LCP's need the use of
tridiagonal matrices. In Haentjens,  Hout and Foulon \cite{hf},
the splitting method of Ikonen and Tovainen is combined with ADI schemes in order
to obtain more efficient numerical results.

In this paper we propose a new approach based both on tree and finite
difference methods. Roughly speaking, our method approximates the CIR
type volatility process through a tree approach already studied in Appolloni, Caramellino and Zanette \cite{bib:acz}, which turns out to be very robust and reliable. And at each step, we make use of a suitable transformation of the asset price process allowing one to take care of a new diffusion process with null correlation w.r.t. the volatility process. Then, by taking into account the conditional behavior with respect to the evolution of the volatility process, we consider a finite difference method to deal with the evolution of the (transformed) underlying asset price process. We note that our procedure can be easily adapted to general stochastic volatility models, in particular to the Hull and White model \cite{HullWhite} and the Stein and Stein model \cite{SteinStein} (see next Remark \ref{noHeston}). We also stress that jumps may be allowed in the dynamics for the underlying asset prices process, but this shall be the subject of a further work.

The paper is organized as follows. In Section \ref{sect-model}, we
introduce the model, we study in details the partial differential equation associated
to the pricing problem (Section \ref{sec_PDE}) and then we set-up our hybrid-finite
difference scheme (Section \ref{sect-tree-fd}). In Section \ref{sect-conv} we give the formal definition of the approximating algorithm and we study the convergence. Section \ref{sect-numerics} is devoted to the numerical results and to comparisons with other existing methods.

\medskip

\medskip \noindent {\bf Acknowledgments.} The authors wish to thank
Alexander Ern and Giuseppe Ruzzi for useful discussions and comments.

\section{Construction of the method}\label{sect-model}
The Heston model \cite{hes} is concerned  with cases where the volatility $V$ is assumed to be stochastic. The dynamics under the risk neutral measure of the share price $S$ and the volatility process $V$ are governed by the stochastic differential equation system
\begin{align*}
&\frac{dS(t)}{S(t)}= (r-\delta)dt+\sqrt{V(t)}\, dZ_S(t),  \\
&dV(t)= \kappa(\theta-V(t))dt+\sigma\sqrt{V(t)}\,dZ_V(t),
\end{align*}
with $S(0)=S_0>0$ and $V(0)=V_0>0$, where $Z_S$ and $Z_V$ are Brownian motions with correlation coefficient $\rho$: $d\<Z_S, Z_V\>(t)=\rho\,dt$. Here $r$ is the risk free rate of interest and $\delta$ the continuous dividend rate.
We assume $\kappa,\theta>0$ and we recall that the dynamics of $V$ follows a CIR process with mean reversion rate $\kappa$ and long run variance $\theta$. The parameter $\sigma$ is called
the volatility of the volatility.

From now on we set
$$
\bar\rho=\sqrt{1-\rho^2}\quad\mbox{and}\quad Z_V=W,\quad Z_S=\rho W+\bar \rho Z,
$$
in which $(W,Z)$ denotes a standard $2$-dimensional Brownian motion. So, the dynamics can be written as
\begin{align}
\label{St}
&\frac{dS(t)}{S(t)}= (r-\delta)dt+\sqrt{V(t)}\,(\rho dW(t)+\bar\rho dZ(t)), \\
\label{vt}
&dV(t)= \kappa(\theta-V(t))dt+\sigma\sqrt{V(t)}\, dW(t).
\end{align}

\smallskip

We consider the diffusion pair $(Y,V)$, where
\begin{equation}\label{Y}
Y_t=\log S_t-\frac \rho{\sigma}\,V_t.
\end{equation}
One has
\begin{align}
\label{Yt1}
&dY(t)= \Big(r-\delta-\frac 12 V_t-\frac \rho{\sigma}\kappa(\theta-V_t)\Big)dt+\bar\rho\,\sqrt{V(t)}\, dZ(t), \\
\label{vt1}
&dV(t)= \kappa(\theta-V(t))dt+\sigma\sqrt{V(t)}\,dW(t),
\end{align}
(recall that $W$ and $Z$ are independent Brownian motions), with
\begin{equation}\label{y0}
Y_0=\log S_0-\frac{\rho}{\sigma}V_0.
\end{equation}
In the following, we define $\mu_Y$ and $\mu_V$ to be the drift coefficient of $Y_t$ and $V_t$ respectively, i.e.
\begin{equation}\label{drifts}
\mu_Y(v)=r-\delta-\frac 12 v-\frac \rho{\sigma}\kappa(\theta-v)
\quad\mbox{and}\quad
\mu_V(v)=\kappa(\theta-v).
\end{equation}
This means that any functional of the pair $(S_t,V_t)$ can be written as a suitable functional of the pair $(Y_t,V_t)$ by using the transformation (\ref{Y}), so $(Y_t,V_t)$ will be our underlying process of interest.

\begin{remark}\label{noHeston}
We shall see in a moment that our numerical procedure is strongly based on the fact that the process $V$ can be efficiently approximated by a tree method and the process $Y$ (to which we shall apply a finite difference method) is an It\^o's process whose coefficients depend on $V$ only and whose driving Brownian motion is independent of the Brownian noise in the stochastic differential equation for $V$, as in \eqref{Yt1}-\eqref{vt1}. It is worth observing that this situation  is standard in all the well-known and studied stochastic volatility models via diffusion processes. In fact, consider a general stochastic volatility model, that is
\begin{align}
\label{St-gen}
&\frac{dS_t}{S_t}= (r-\delta)dt+\eta(V_t)\,(\rho dW_t+\bar\rho dZ_t), \\
\label{vt-gen}
&dV_t= \mu_V(V_t)dt+\sigma_V(V_t)\, dW_t,
\end{align}
where $\eta$, $\mu_V$ and $\sigma_V$ are suitable functions. We assume that there exist
two functions $G$ and $\xi$ such that
\begin{equation}\label{cond}
\int_0^t \eta(V_s)dW_s=G(V_t)+\int_0^t\xi(V_s)ds.
\end{equation}
Note that if one requires that $G$ is twice differentiable, by the It\^o's formula \eqref{cond} is equivalent to the conditions
\begin{equation}\label{cond-bis}
\eta(v)=G'(v)\sigma_V(v)\quad\mbox{and}\quad \xi(v)=-G'(v)\mu_V(v)-\frac 12 \,G''(v)\sigma_V(v)^2.
\end{equation}
Set now
$$
Y_t=\ln S_t-\rho G(V_t),
$$
so that $(S_t,V_t)=(e^{Y_t+\rho G(V_t)},V_t)$. Then, by applying the It\^o's formula, one immediately gets
\begin{align*}
&dY_t= \mu_Y(V_t)dt+\bar\rho \eta(V_t) dZ_t, \\
&dV_t= \mu_V(V_t)dt+\sigma_V(V_t)\, dW_t,
\end{align*}
in which $\mu_Y(v)=r-\delta-\frac 12 \eta(v)^2-\rho\xi(v)$. So, we are in a situation similar to that described for the Heston model, in which  $G(v)=\frac v\sigma$ and $\xi(v)=-\frac 1\sigma \mu_V(v)$.
But such functions $G$ and $\xi$ actually exist also in the other well-known stochastic volatility models in the literature, that is:
\begin{itemize}
\item Hull and White model \cite{HullWhite}: equations \eqref{St-gen} - \eqref{vt-gen} are
$$
\frac{dS_t}{S_t}= (r-\delta)dt+\sqrt{V_t}\,(\rho dW_t+\bar\rho dZ_t)\quad\mbox{and}\quad
dV_t= \mu V_tdt+\sigma V_t\, dW_t
$$
and here, $G(v)=\frac 2\sigma \sqrt v$ and $\xi(v)=\big(-\frac\mu\sigma+\frac\sigma 4\big)\sqrt v$ - the fact that $G$ is a square root is not really a problem because $V$ is a geometric Brownian motion, so we can restrict to the half-space $v>0$ and the It\^o's formula applies as well;

\item Stein and Stein model \cite{SteinStein}: equations \eqref{St-gen} - \eqref{vt-gen} are
$$
\frac{dS_t}{S_t}= (r-\delta)dt+V_t\,(\rho dW_t+\bar\rho dZ_t)\quad\mbox{and}\quad
dV_t= \kappa(\theta-V_t)dt+\sigma \, dW_t
$$
and here, $G(v)=\frac {v^2}{2\sigma}$ and $\xi(v)=-\frac1\sigma\kappa v(\theta-v)+\frac 12\sigma$.
\end{itemize}

\end{remark}

\subsection{The associated pricing PDE in a small time interval}\label{sec_PDE}

Let $f=f(y,v)$ be a function of the time and the space-variable pair $(y,v)$. For $h$ small, we need to compute (an estimate for) the quantity $u(t,y,v)$ defined through
$$
u(t,y,v)=\E\big(f(Y_{t+h}^{t,y,v},V_{t+h}^{t,v})\big),
$$
in which $(Y^{t,y,v},V^{t,v})$ denotes the solution to (\ref{Yt1}) and (\ref{vt1}) with the starting condition $(Y_t,V_t)$ $=(y,v)$. In our mind, the time instant $t$ plays the role of a dicretization instant in $[0,T]$, that is $t=nh$, so $t+h=(n+1)h$ stands for the next discretizing time. It is well-known that, for a suitable function $f$, $u$ solves the Heston parabolic partial differential equation. We do not use this fact but let us point out that results for such kind of degenerate parabolic equations have been recently developed  in \cite{FeehanPop1} and \cite{FeehanPop2}.

To our purposes, we first notice that
$$
\E\big(f(Y_{t+h}^{t,y,v},V_{t+h}^{t,v})\big)
=\E\Big(\E\big(f(Y_{t+h}^{t,y,v},V_{t+h}^{t,v})\,|\,\cl F^W_{t+h}\big)\Big)
$$
where $\cl F^W_{t+h}=\sigma(W_u\,:\,u\leq t+h)$. But conditional on $\cl F^W_{t+h}$, the volatility process $V$ can be considered deterministic and the process $Y$ turns out to have constant coefficients. More precisely, for $g\in L^2([t,t+h])$ with $g\geq 0$ a.e. and $g_t=v$, set
\begin{equation}\label{Ug}
U^{(g),t,y}_{t+h}= y+\int_t^{t+h}\mu_Y(g_s)ds+\bar\rho\,\int_t^{t+h}\sqrt{g_s}\, dZ(s).
\end{equation}
Then
$$
\E\big(f(Y_{t+h}^{t,y,v},V_{t+h}^{t,v})\,|\,\cl F^W_{t+h}\big)
=\E\big(f(U^{(g),t,y}_{t+h},g_{t+h}\big)|_{g=V^{t,v}}.
$$
We define now
$$
\bar u(t,y;g)=\E\big(f(U^{(g),t,y}_{t+h},g_{t+h})\big),
$$
so that
$$
\E\big(f(Y_{t+h}^{s,y,v},V_{t+h}^{s,v})\,|\,\cl F^W_{t+h}\big)
=\bar u(t,y;V^{t,v})
$$
and therefore,
\begin{equation}\label{key}
u(t,y,v)=\E\big(\bar u(t,y;V^{t,v})\big).
\end{equation}
Let us now discuss the quantity $\bar u(t,y;g)$,  for $g$ fixed  as required above, that is
\begin{equation}\label{g}
\mbox{$g\,:\,[t,t+h]\to \R_+$ is continuous and $g_t=v$}.
\end{equation}
Set $ U^{(g),s,y}_{t+h}$ as the solution $U^{(g)}$ at time $t+h$, with starting condition $ U^{(g)}_s=y$ of the following stochastic differential equation \emph{with deterministic (although path dependent) coefficients}:
\begin{equation}\label{Ug1}
dU^{(g)}_{u}=\mu_Y(g_u)du+\bar\rho\,\sqrt{g_u}\,dZ_u.
\end{equation}
Recall that the associated infinitesimal generator is given by
\begin{equation}\label{LbarUg}
L^{(g)}_u=\mu_Y(g_u)\partial_y+\frac {1}2\,\bar\rho^2g_u\partial^2_{yy}.
\end{equation}
So, we get
$$
\bar u(t,y;g)=\bar u(s,y;g)|_{s=t},\quad
\mbox{with}\quad
\bar u(s,y;g)=\E\big(f(U^{(g),s,y}_{t+h};g_{t+h})\big).
$$
Now, from the Feynman-Kac formula, the function $(s,y)\mapsto \bar u(s,y;g)$ solves the parabolic PDE Cauchy problem
\begin{equation}\label{PDE-barug}
\begin{array}{ll}
\displaystyle
\partial_s \bar u(s,y;g)+L^{(g)}_s\bar u(s,y;g)=0 & y\in\R, s\in (t,t+h),\smallskip\\
\displaystyle
\bar u(t+h,y;g)
=f(y,g_{t+h}) & y\in \R.
\end{array}
\end{equation}
To be formally correct, one should precise the right conditions on $f$ in order to get the solution of \eqref{PDE-barug}. We do not enter in these arguments because we aim to give here only the main ideas that inspired the construction of our numerical procedure.

Once the problem (\ref{PDE-barug}) is solved, we can proceed to compute $u(t,y,v)$ by using (\ref{key}). We stress that the fixed path $g$ plays the role of a parameter and the solution to
(\ref{PDE-barug}) depends in general on the whole trajectory of $g$.

\smallskip

We consider now the case $h\simeq 0$, so that, by (\ref{g}),   $g_s\simeq g_t=v$ and $\mu_Y(g_s)\simeq \mu_Y(g_t)=\mu_Y(v)$. This numerically brings to replace (\ref{PDE-barug}) with a PDE problem \emph{with constant coefficients}. More precisely, we consider the approximation $\hat u^h(s,y;v,g_{t+h})$ for $\bar u(t,y;g)$ given by the solution to
\begin{equation}\label{PDE-barug-h}
\begin{array}{ll}
\displaystyle
\partial_s \hat u^h(s,y;v,g_{t+h})+L^{(v)}\hat u^h(s,y;v,g_{t+h})=0 & y\in\R, s\in (t,t+h),\smallskip\\
\displaystyle
\hat u^h(t+h,y;v,g_{t+h})
=f(y,g_{t+h}) & y\in \R,
\end{array}
\end{equation}
with
$$
L^{(v)}=\mu_Y(v)\partial_y +\frac 12\bar\rho^2 v\partial^2_{yy}.
$$
Let us remark that the solution to (\ref{PDE-barug-h}) actually depends on $g$ only through $v=g_t$ (appearing in the coefficients of the second order operator) and $g_{t+h}$ (appearing in the Cauchy condition), that is why we used the notation $\hat u^h(s,y;v,g_{t+h})$. In contrast, the function solving (\ref{PDE-barug}) depends in principle  on the whole trajectory $g$ over the time interval $[t,t+h]$.

Now, problem \eqref {PDE-barug-h} can be easily solved by using a finite difference numerical method. Numerical reasonings suggest the use of an implicit approximation (in time) if $v$ is ``far enough'' from zero, otherwise an explicit method should be considered - details are given in Section \ref{sect-implicit} and \ref{sect-explicit}. This means that one fixes a space-step $\Dy_h$ and a space-grid $\mathcal{Y}^h=\{y_j=Y_0+j\Dy_h\}_{j\in\Z}$
splitting the real line and approximates the solution $\hat u^h(s,y;v,g_{t+h})$ to \eqref{PDE-barug-h} on the grid $\mathcal{Y}^h$ by means of a linear operator (infinite dimensional matrix) $\Pi^h(v)=(\Pi^h(v)_{i,j})_{i,j\in\Z}$. In other words, one gets
$$
\hat u^h(s,y_i;v,g_{t+h})\simeq
\sum_{j\in\Z}\Pi^h(v)_{i,j}f(y_j,g_{t+h}),\quad i\in\Z.
$$
Now, recalling (\ref{key}) and the fact that $\bar u^h\simeq \hat u^h$, on the grid $\mathcal{Y}$ the function $u$ is approximated through
\begin{equation}\label{key-approx}
u(t,y_i,v)\simeq
\E\big(\hat u^h(t,y_i;v,V^{t,v}_{t+h})\big)
\simeq \sum_{j\in\Z}\Pi^h(v)_{i,j}\E\big(f(y_j,V^{t,v}_{t+h})\big),\quad i\in\Z .
\end{equation}
We stress that the expectation on the r.h.s. above is now written in terms of the process $V$ only, and this is the key point of our story because we can now use the tree method in \cite{bib:acz}. But we will examine in depth this point in a moment.

In practice, one cannot solve the PDE problem over the whole real line. So, one takes a positive integer $M_h>0$ such that $M_h\Dy_h\to+\infty$ as $h\to0$ and considers a discretization of the (space) interval $[-M_h\Dy_h+Y_0,Y_0+M_h\Dy_h]$ in $2M_h+1$ equally spaced points $y_{j}=Y_0+j\Dy_h$, $j\in \mathcal{J}_{M_h}=\{-M_h,\ldots,M_h\}$. Then, the grid $\mathcal{Y}^h_{M_h}=\{y_j=Y_0+j\Dy_h\}_{j\in\mathcal{J}_{M_h}}$ is finite and the approximation of $\hat u^h_{n}(nh, y; v,g_{t+h})$ is done by adding to \eqref{PDE-barug-h} suitable boundary conditions. By calling again $\Pi^h(v)$ the matrix (now, finite dimensional) giving the solution from the finite difference approach, we still obtain
\begin{equation}\label{key-approx-finite}
\E\big(f(Y_{t+h}^{t,y,v},V_{t+h}^{t,v})\big)\Big|_{y=y_i}\simeq \sum_{j\in\mathcal{J}_{M_h}}\Pi^h(v)_{i,j}\E\big(f(y_j,V^{t,v}_{t+h})\big),\quad i\in\mathcal{J}_{M_h}.
\end{equation}

\subsection{The hybrid tree-finite difference approach}\label{sect-tree-fd}
We describe the main ideas of our approximating algorithm by means of an example, which was our starting point.

Consider an American option with maturity $T$ and payoff function $(\Phi(S_t))_{t\in[0,T]}$. First of all, by using (\ref{Y}) we replace the pair $(S,V)$ with the pair $(Y,V)$, so the obstacle will be given by
$$
\Psi(Y_t,V_t)=\Phi(e^{Y_t+\frac\rho{\sigma}V_t}),\quad t\in[0,T].
$$
The price at time $0$ of such an option
is then approximated by a backward dynamic programming algorithm, working as follows. First, consider a discretization of the time interval $[0,T]$ into $N$ subintervals of length $h=T/N$: $[0,T]=\cup_{n=0}^{N-1}[nh,(n+1)h]$. Then the price $P(0,Y_0,V_0)$ of such an American option is numerically approximated through the quantity $P_h(0,Y_0,V_0)$ which is iteratively defined as follows: for $(y,v)\in\R\times\R_+$,
\begin{equation*}\label{backward}
  \begin{cases}
    P_h(T,y,v)= \Psi(y,v)\quad \mbox{and as $n=N-1,\ldots,0$}\\
   P_h(nh,y,v) =  \max \Big\{\Psi(y,v), e^{-rh}
   \E\Big(P_h\big((n+1)h, Y_{(n+1)h}^{nh,y,v}, V_{(n+1)h}^{nh,v}\big)
   \Big)\Big\}.
  \end{cases}
\end{equation*}
From the financial point of view, this means to allow the exercise at the fixed times $nh$, $n=0,\ldots,N$. Now, what we are going to set up is a mixing of a tree method for the process $V$ and  a finite difference method to handle the noise in $Y$ (which is independent of the noise driving $V$). In fact,  the expectations appearing in the backward induction can be written as expectations of functions of the process $V$ only,  such functions being solution to parabolic PDE's.
So, we proceed as described in the previous section:  we fix a grid on the $y$-axis $\mathcal{Y}^h_{M_h}=\{y_j=Y_0+j\Dy_h\}_{j\in\mathcal{J}_{M_h}}$, with $\mathcal{J}_{M_h}=\{-M_h,\ldots,M_h\}$,  and we approximate the above conditional expectations for $y=y_i\in\mathcal{Y}^h_{M_h}$ by using the matrix $\Pi^h(v)$ from the finite difference method. So, as already seen in \eqref{key-approx-finite}, we write
\begin{equation}\label{key-approx-bis}
\E\Big(P_h\big((n+1)h, Y_{(n+1)h}^{nh,y,v}, V_{(n+1)h}^{nh,v}\big)\Big)\Big|_{y=y_i}\simeq
\sum_{j\in\mathcal{J}_{M_h}}\Pi^h(v)_{i,j}\E\big(P_h\big((n+1)h, y_j, V_{(n+1)h}^{nh,v}\big),\quad i\in\mathcal{J}_{M_h}.
\end{equation}
By resuming, the price $P(0,y,v)$ in $y\in\mathcal{Y}^h_{M_h}$ can be numerically computed from the function $\hat P_h(0,y,v)$ defined on the grid $\mathcal{Y}^h_{M_h}$ as follows:
\begin{equation}\label{backward-bis}
\left\{  \begin{array}{l}
    \hat P_h(T,y_i,v)= \Psi(y_i,v), \quad \mbox{$i\in\mathcal{J}_{M_h}$ and as $n=N-1,\ldots,0$},\\
   \hat P_h(nh,y_i,v) = \max \Big\{\Psi(y_i,v), e^{-rh}
\displaystyle\sum_{j\in\mathcal{J}_{M_h}}\Pi^h(v)_{i,j}\E\big(\hat P_h\big((n+1)h, y_j, V_{(n+1)h}^{nh,v}\big)   \big)\Big\},\ i\in\mathcal{J}_{M_h}.
  \end{array}\right.
\end{equation}
We stress that the backward induction (\ref{backward-bis}) is now written in terms of the process $V$ only,
and here the binomial tree method in \cite{bib:acz} comes on, we see how in a moment. First,  let us briefly recall the binomial tree procedure in \cite{bib:acz}.

\smallskip

For $n=0,1,\ldots,N$, consider the lattice
\begin{equation}\label{state-space-V}
\mathcal{V}_n^h=\{v_{n,k}\}_{k=0,1,\ldots,n}\quad\mbox{with}\quad
v_{n,k}=\Big(\sqrt {V_0}+\frac\sigma 2(2k-n)\sqrt{h}\Big)^2\I_{\sqrt {V_0}+\frac\sigma 2(2k-n)\sqrt{h}>0}
\end{equation}
(notice that $v_{0,0}=V_0$) and for each fixed $v_{n,k}\in\mathcal{V}_n^h$, we define
\begin{align}
\label{kd}
&k_d^h(n,k) =\max\{k^*\,:\, 0\leq k^*\leq k \mbox{ and }v_{n,k}+\mu_V(v_{n,k})h \ge v_{n+1, k^*}\},\\
\label{ku}
&k_u^h(n,k) =\min\{k^*\,:\, k+1\leq k^*\leq n+1\mbox{ and }v_{n,k}+\mu_V(v_{n,k})h \le v_{n+1, k^*}\}
\end{align}
with the understanding $k_d^h(n,k)=0$ if $\{k^*\,:\, 0\leq k^*\leq k \mbox{ and }v_{n,k}+\mu_V(v_{n,k})h \ge v_{n+1, k^*}\}=\emptyset$ and $k_u^h(n,k)=n+1$ if $\{k^*\,:\, k+1\leq k^*\leq n+1\mbox{ and }v_{n,k}+\mu_V(v_{n,k})h \le v_{n+1, k^*}\}=\emptyset$.
The transition probabilities are defined as follows: starting from the node $(n,k)$ the probability that the process jumps to $k_u^h(n,k)$ at time-step $n+1$ is set as
\begin{equation}\label{pik}
p^h_{k_u^h(n,k)}
=0\vee \frac{\mu_V(v_{n,k})h+ v_{n,k}-v_{n+1,k_d^h(n,k)} }{v_{n+1,k_u^h(n,k)}-v_{n+1,k_d^h(n,k)}}\wedge 1.
\end{equation}
And of course, the jump to $(n+1,k_d^h(n,k))$ happens with probability $p^h_{k_d^h(n,k)}=1-p^h_{k_u^h(n,k)}$.
This gives rise to a Markov chain $(\bar V^h_{n})_{n=0,\ldots,N}$ that weakly converges, as $h\to 0$, to the diffusion process $(V_{t})_{t\in[0,T]}$ and turns out to be a robust tree approximation for the CIR process $V$. This means that we can approximate the expectation of suitable  functionals of the diffusion $V$ with the same expectation evaluated on the Markov chain $\bar V^h$. In particular, for a function $g$ we write
$$
\E\big(g(V_{(n+1)h}^{nh,v_{n,k}})\big)\simeq
\E\big(g(\bar V^h_{n+1})\,|\,\bar V^h_{n}=v_{n,k}\big)
=g(v_{n+1,k_u^h(n,k)})p^h_{k_u^h(n,k)}+g(v_{n+1,k^h_d(n,k)})p^h_{k_d^h(n,k)}.
$$
So, at step $n$, we can approximate the expectation in the backward induction (\ref{backward-bis}) on the lattice $\mathcal{V}_n^h$ as
$$
\E\big(\hat P_h\big((n+1)h, y_j, V_{(n+1)h}^{nh,v}\big)\big)\Big|_{v=v_{n,k}}
\!\!\!\simeq \sum_{k^*\in\{k_u^h(n,k),k_d^h(n,k)\}}\hat P_h\big((n+1)h, y_j, v_{n+1,k^*}\big)p^h_{k^*}.
$$
We can finally write the backward induction giving our approximating algorithm: for $n=0,1,\ldots,N$, we define $\tilde P_h(nh,y,v)$ for $(y,v)\in \mathcal{Y}^h_{M_h}\times\mathcal{V}_n^h$ by
\begin{equation}\label{backward-ter}
  \begin{cases}
    \tilde P_h(T,y_i,v_{N,k})= \Psi(y_i,v_{N,k})\quad \mbox{$i\in\mathcal{J}_{M_h}$ and $v_{N,k}\in\mathcal{V}_n^h$, and as $n=N-1,\ldots,0$}\\
\displaystyle   \tilde P_h(nh,y_i,v_{n,k}) =  \max \Big\{\Psi(y_i,v), e^{-rh}
\sum_{k^*\!,\,j}
\Pi^h(v_{n,k})_{i,j}\tilde P_h\big((n+1)h, y_j, v_{n+1,k^*}\big)p^h_{k^*}\Big\},\\
\qquad\qquad\qquad\qquad i\in\mathcal{J}_{M_h}\quad\mbox{and}\quad v_{n,k}\in\mathcal{V}_n^h,
  \end{cases}
\end{equation}
where the sum above is done for $k^*\in\{k_u^h(n,k),k_d^h(n,k)\}$ and $j\in\mathcal{J}_{M_h}$.
Notice that, at time step  $n$, for every fixed $i\in\mathcal{J}_{M_h}$ and $k=0,\ldots,n$   the sum in \eqref{backward-ter} can be seen as an integral w.r.t. the measure
\begin{equation}\label{mu-h}
\mu^h(y_i,v_{n,k};A\times B)
=\sum_{k^*\in\{k_u^h(n,k),k_d^h(n,k)\}}\sum_{j\in\mathcal{J}_{M_h}}
\Pi^h(v_{n,k})_{i,j}p^h_{k^*}\delta_{\{y_j\}}(A)\delta_{\{v_{n+1,k^*}\}}(B)
\end{equation}
for every Borel sets $A$ and $B$, $\delta_{\{a\}}$ denoting the Dirac mass in $a$, so that $\mu^h(y_i,v_{n,k};\cdot)$ is a discrete measure on $\mathcal{Y}^h_{M_h}\times \mathcal{V}_{n+1}^h$.

Now, in next Section \ref{sect-conv} we shall be able to prove that, for small values of  $h$, $\Pi^h(v)$ is a stochastic matrix. This gives that $\mu^h(y_i,v_{n,k};\cdot)$ is actually a probability measure, that can be interpreted as a transition probability measure. Thus, we are doing expectations on a Markov chain $(\bar Y^h_n,\bar V^h_n)_{n=0,1,\ldots,N}$, whose state-space, at time step $n$, is given by $\mathcal{Y}^h_{M_h}\times \mathcal{V}_{n}^h$ and whose transition probability measure at time step  $n$ is given by $\mu^h(y_i,v_{n,k};\cdot)$ in \eqref{mu-h}. Moreover, we shall prove that, under appropriate conditions on $\Dy_h$ and $M_h$ such that, as $h\to 0$,  $\Dy_h\to 0$ and $M_h\Dy_h\to\infty$ (see next \eqref{dep-h} and \eqref{dep-par}), the family of Markov chains $(\bar Y^h,\bar V^h)_h$ weakly converges to the diffusion process $(Y,V)$ . And this gives the convergence of our hybrid tree-finite difference algorithm approximating the Heston model. We shall finally discuss the convergence of the backward induction \eqref{backward-ter} to the price of the associated American option written on the Heston model.

\section{The convergence of the algorithm}\label{sect-conv}

We first set up the finite difference method we take into account. Then, in Section \ref{sect-convMC}, we formally define the approximating Markov chain and prove the weak convergence to the Heston model in the path space.

For ease of notation, for a while we drop the dependence on the time-step $h$ for the space-step $\Dy_h$ and the number $M_h$ related to the points of the space-grid, so we simply write $\Dy$ and $M$, as well as $\mathcal{J}_M=\{-M,\ldots,M\}$ and $\mathcal{Y}_M=\{y_i=Y_0+i\Dy\}_{i\in\mathcal{J}_M}$.

\subsection{The finite difference scheme for the PDE problem \eqref{PDE-barug-h}}
As described in Section \ref{sect-tree-fd}, at each time step $n$ we need to numerically solve \eqref{PDE-barug-h} for $t=t_n=nh$. So, we briefly describe the finite difference method we apply to problem \eqref{PDE-barug-h}, outlining some 	
key properties of the associated operator allowing us to prove the convergence. For further information on finite difference methods for partial differential equations we refer for instance to \cite{strikwerda}.

\smallskip

Let $t=nh$, $v$ and $v^*=g_{nh+h}$ be fixed and let us set $u^n_j=\tilde u^h_{n}(nh, y_j)$ the discrete solution of \eqref{PDE-barug-h} at time $nh$ on the point $y_j$ of the grid $\mathcal{Y}_{M}$ - for simplicity of notations, we do not stress on $u^n_j$ the dependence on $v$ (from the coefficients of the PDE), $v^*$ (from the Cauchy problem) and $h$.

It is well known that the behavior of the solution to \eqref{PDE-barug-h} changes with respect the magnitude of the rate between the diffusion coefficient ($\rho^2 v/2$) and the advection term ($\mu_Y(v)$). To deal with these cases, we fix a small real threshold $\epsilon>0$ and in the following we shall describe how to solve both the case $v<\epsilon$ and $v>\epsilon$ by applying an explicit in time and an implicit in time approximation respectively.

It is well known that for a big enough diffusion coefficient, to avoid over-restrictive conditions on the grid steps, it is suggested to apply implicit finite differences to problem \eqref{PDE-barug-h}. In this case,
the discrete solution $\{u^n_j\}_{j\in\Z}$ at time $nh$ will then be computed in terms of the solution $\{u^{n+1}_j\}_{j\in\Z}$ at time $(n+1)h$ by solving the following discrete problem:
\begin{equation}\label{discr_eq}
\begin{array}{l}
\Frac{u^{n+1}_j-u^n_j}{h}+\mu_Y(v)\Frac{u^{n}_{j+1}-u^{n}_{j-1}}{2\Dy}+\frac{1}{2}\bar\rho^2\ v\ \Frac{u^{n}_{j+1}-2u^n_j+u^{n}_{j-1}}{\Dy^2}=0, \quad j\in\Z
\end{array}
\end{equation}
where $\Delta y=y_j-y_{j-1}$, $\forall j\in\Z$.

On the other hand, when the diffusion coefficient is small compared with the reaction one, it is suggested to apply an explicit in time approximation coupled with a forward or backward finite differences for the first order term $u_x$ depending on the sign of the reaction coefficient.

Specifically, for $v$ close to 0, that is $v<\epsilon$, we solve the problem by the following approximation schemes: when $\mu_Y(v)>0$,
\begin{equation}\label{discr_eq_expl_p}
\begin{array}{l}
\Frac{u^{n+1}_j-u^n_j}{h}+\mu_Y(v)\Frac{u^{n+1}_{j+1}-u^{n+1}_{j}}{\Dy}+\frac{1}{2}\bar\rho^2\ v\ \Frac{u^{n+1}_{j+1}-2u^{n+1}_j+u^{n+1}_{j-1}}{\Dy^2}=0, \quad j\in\Z,
\end{array}
\end{equation}
while, when $\mu_Y(v)<0$,
\begin{equation}\label{discr_eq_expl_m}
\begin{array}{l}
\Frac{u^{n+1}_j-u^n_j}{h}+\mu_Y(v)\Frac{u^{n+1}_{j}-u^{n+1}_{j-1}}{\Dy}+\frac{1}{2}\bar\rho^2\ v\ \Frac{u^{n+1}_{j+1}-2u^{n+1}_j+u^{n+1}_{j-1}}{\Dy^2}=0, \quad j\in\Z.
\end{array}
\end{equation}
As previously mentioned at the end of the Section \ref{sec_PDE}, for the numerical tests one has to deal with a finite grid $\mathcal{Y}_M^h=\{y_j\}_{j\in\mathcal{J}_{M}}$  and problems \eqref{discr_eq} and \eqref{discr_eq_expl_p} have to be coupled with suitable numerical boundary conditions.
Here, we assume that the boundary of the domain values are defined by the following relations (Neumann-type boundary conditions): in the implicit case we set
\begin{equation}
u^{n}_{-M-1}=u^{n}_{-M+1},
\quad u^{n}_{M+1}=u^{n}_{M-1}\label{num_boundary_cond_impl},
\end{equation}
whereas in the explicit case we set
\begin{equation}
u^{n+1}_{-M-1}=u^{n+1}_{-M+1},
\quad u^{n+1}_{M+1}=u^{n+1}_{M-1}\label{num_boundary_cond_expl}
\end{equation}
Other conditions can surely be selected, for example the two \emph{boundary} values $u^{n+1}_{-M}$ and $u^{n+1}_{M}$ may be a priori fixed by a known constant (this typically appears in financial problems). Specifically, other types of boundary conditions can be handled by using Proposition \ref{prop_boundary_stima} (see Appendix \ref{app-boundary}) and all arguments that follow apply as well (for details, see next Remark \ref{boundary-cond}).
So, hereafter we set
\begin{equation}\label{alphabeta}
\alpha=\frac{h}{2\Dy}\,\mu_Y(v)\quad\mbox{and}\quad\beta=\frac{h}{2\Dy^2}\,\bar\rho^2v.
\end{equation}

\subsubsection{The case $v>\epsilon$}\label{sect-implicit}
By applying implicit finite differences \eqref{discr_eq} coupled with boundary conditions \eqref{num_boundary_cond_impl}, we get the solution $u^n=(u^n_{-M},\ldots,u^n_M)^T$ by solving the following linear system
\begin{equation}\label{lin_sys}
A\, u^n = u^{n+1},
\end{equation}
where $A$ is the $(2M+1)\times (2M+1)$ tridiagonal real matrix given by
\begin{equation}\label{matrixA}
A = \left(
\begin{array}{ccccc}
1+2\beta & -2\beta & & & \\
\alpha-\beta & 1+2\beta & -\alpha-\beta & & \\
 & \ddots & \ddots & \ddots &  \\
 &   & \alpha-\beta & 1+2\beta & -\alpha-\beta \\
 &   &        &   -2\beta & 1+2\beta
\end{array}
\right).
\end{equation}

We immediately prove that the solution $u^n$ to \eqref{lin_sys} actually exists at least when $\beta>|\alpha|$ (we will see later that this is not a restrictive condition).

\begin{proposition}\label{propr_A}
Assume that $\beta>|\alpha|$. Then $A$ is invertible and $A^{-1}$ is a stochastic matrix, that is all entries are non negative and, for $\Vunita = (1,...,1)^T$,
$A\Vunita=\Vunita$.
\end{proposition}

\textbf{Proof.} The matrix $A=(a_{ij})_{i,j\in\mathcal{J}_M}$ satisfies

\medskip

$\quad$
(P1) $A\Vunita=\Vunita$, i.e. $\sum_{j=-M}^M a_{ij}=1$ for $i\in\mathcal{J}_M$

\medskip

and for $\beta>|\alpha|$, one has also

\medskip

$\quad$
(P2) $a_{ii}>0$ for $i\in\mathcal{J}_M$ and for $j\in\mathcal{J}_M$, $j\ne i$, $a_{ij}\leq 0$,

\smallskip

$\quad$
(P3) $A$ is strict or irreducibly diagonally dominant, i.e. $\sum_{j\in\mathcal{J}_M,j\ne i} |a_{ij}|< a_{ii}$ for $i\in\mathcal{J}_M$.

\medskip

(P2)-(P3) give that $A$ is an invertible $M$-matrix (see for instance \cite{mmatrix}), so that $A^{-1}$ is non-negative (i.e. $a^{-1}_{ij}\geq 0$, $i,j=-M,...,M$).
Moreover, by (P1), $\Vunita = A^{-1} \Vunita $.
\cvd

For each $l\in \N$ and $y_i\in\mathcal{Y}_M$, we consider the polynomial $(y-y_i)^l$ and we call $\psi^i_l(y)\in\R^{2M+1}$ the associated (vector) function of  $y\in\mathcal{Y}_M$:
\begin{equation}\label{psi}
\big(\psi^i_l(y)\big)_k =
(y_k-y_i)^l=\Dy^l(k-i)^l, \quad k\in\mathcal{J}_M.
\end{equation}
In next Section \ref{sect-convMC} we need to deal with $A^{-1}\psi^i_l(y)$ for $l\leq 4$ and $i\in\mathcal{J}_M$. So, we study such objects.
By Proposition \ref{propr_A}, for $\beta>|\alpha|$ one has that $A$ is invertible and we may then compute $A^{-1}\psi^i_l(y)$. We also notice that $\psi^i_0(y)=\Vunita$, so that $A^{-1}\psi^i_0(y)=A^{-1}\Vunita=\Vunita$.

\smallskip

In the following, the symbol $[\cdot]$ will stand for the floor function and we use the understanding $ \sum_{k=1}^0 (\cdot)_k := 0$. Moreover, we let
$\e_i$ denote the standard orthonormal basis: for  $i,k\in\mathcal{J}_M$, $(\e_i)_k=0$ for $k\neq i$ and $(\e_i)_k=1$ if $k=i$.

\begin{lemma}\label{Apsi}
Let $\psi^i_l(y)$ be defined in \eqref{psi}. Then for every $l\in\N$ and $i\in\mathcal{J}_M$ one has
\begin{equation}\label{Apsi_form}
A\psi^i_l(y)=\psi^i_l(y)-\sum_{j=0}^{l-1}\coeffbin{l}{j}a_{l-j}\Dy^{l-j}\psi^i_j(y)+b^{-M}_{l,i}
\e_{-M}+b^M_{l,i}\e_M,
\end{equation}
where
\begin{equation}\label{an}
a_n=(\beta-\alpha)(-1)^n+(\beta+\alpha),\quad n\in\N,
\end{equation}
that is $a_n=2\beta$ if $n$ is even and $a_n=2\alpha$ if $n$ is odd, and the coefficients $b^{\pm M}_{l,i}$ are given by
\begin{equation}\label{B}
b^{\pm M}_{l,i}=\pm 2\displaystyle\sum_{j=0,l-j\ odd}^{l-1}\coeffbin{l}{j}(\beta\pm\alpha)\Dy^{l-j}(y_{\pm M}-y_i)^j.
\end{equation}
Moreover, $b^{\pm M}_{l,i}$ can be bounded as follows:
\begin{equation}\label{stima-B}
|b^{\pm M}_{l,i}|\leq 2(\beta\pm\alpha)(\Dy+|y_{\pm M}-y_i|)^{l}
\end{equation}
\end{lemma}

\textbf{Proof.}
For $k\in\mathcal{J}_M$ with $k\neq \pm M$, we have
\begin{align*}
\big(A\psi^i_l(y)\big)_k
&=-(\beta-\alpha)\big(\psi^i_l(y)\big)_{k-1}
+(1+2\beta)\big(\psi^i_l(y)\big)_{k}
-(\beta+\alpha)\big(\psi^i_l(y)\big)_{k+1}\\
&=-(\beta-\alpha)(y_k-y_i-\Dy)^l
+(1+2\beta)(y_k-y_i)^l
-(\beta+\alpha)(y_k-y_i+\Dy)^l\\
&=(1+2\beta)(y_k-y_i)^l
-\sum_{j=0}^l\coeffbin{l}{j}
(y_k-y_i)^{j} \big((\beta-\alpha)(-1)^{l-j}
+(\beta+\alpha)\big)\Dy^{l-j}\\
&=(y_k-y_i)^l
-\sum_{j=0}^{l-1}\coeffbin{l}{j}
(y_k-y_i)^{j} a_{l-j}\Dy^{l-j}.
\end{align*}
On the other hand, following the same reasonings for $k=\pm M$ we easily obtain
$$
\big(A\psi^i_l(y)\big)_{\pm M} = (y_{\pm M}-y_i)^l-2\beta\displaystyle\sum_{j=0}^{l-1}\coeffbin{l}{j}(\mp \Dy)^{l-j}(y_{\pm M}-y_i)^j
$$
and by  using \eqref{B} we get the full form \eqref{Apsi_form}. Finally, the estimate in \eqref{stima-B} immediately follows from the Newton's binomial formula.
\cvd

We are now ready to characterize $A^{-1}\psi^i_l(y)$, for every polynomial $\psi^i_l(y)$ as in \eqref{psi}.

\begin{proposition}\label{prop-psi}
Suppose that $\beta>|\alpha|$ and for $l\geq 1$ let $\gamma_{l, k}$, $k\leq l$, be iteratively (backwardly) defined as follows: $\gamma_{l,l}=a_0$ and
\begin{equation}\label{gamma-lk}
\gamma_{l,k}=
\coeffbin{l}{k}a_{l-k}\Dy^{l-k}+
\sum_{j=k+1}^{l-1}\gamma_{l,j}\coeffbin{j}{k}a_{j-k}
\Dy^{j-k},\quad
k=l-1,\ldots,0,
\end{equation}
where the coefficients $a_n$ are given in \eqref{an}. Then
$$
A^{-1}\psi^i_l(y)=\psi^i_l(y)+\sum_{j=0}^{l-1}\gamma_{l,j}\psi^i_j(y)+\phi^i_{l,M}(y),
$$
in which
\begin{equation}\label{boundary_term}
\phi^i_{l,M}(y)=T^{-M}_{l,i} A^{-1}\e_{-M}+T^{M}_{l,i} A^{-1}\e_{M}\quad\mbox{with}
\quad T^{\pm M}_{l,i}=-b^{\pm M}_{l,i}-\sum_{j=0}^{l-1}\gamma_{l,j}b^{\pm M}_{j,i},
\end{equation}
the $b^{\pm M}_{j,i}$'s being given in \eqref{B}. Moreover, if $\Dy\leq 1$, $M\Dy\geq 1$ and $l2^{l+2}(\beta\Dy^2+|\alpha|\Dy)\leq 1$, the following estimate holds for $T^{\pm M}_{l,i}$: for every $i\in\mathcal{J}_M$,
\begin{equation}\label{est-T}
|T^{\pm M}_{l,i}|\leq
4(\beta\pm\alpha)\Dy^l(1+2M)^{l}.
\end{equation}

\end{proposition}

\proof It is clear that $A^{-1}\psi^i_l(y)=\psi^i_l(y)+\sum_{j=1}^{l-1}\gamma_{l,j}\psi^i_j(y)+\phi^i_l(y)$ if and only if
$$
\psi^i_l(y)=A\psi^i_l(y)+\sum_{j=0}^{l-1}\gamma_{l,j}A\psi^i_j(y)+A\phi^i_{l,M}(y).
$$
We call $(*)$ the r.h.s. above. By using Lemma \ref{Apsi}
and setting
$$ \theta^i_l= b^{-M}_{l,i}\e_{-M}+b^M_{l,i}\e_M,$$
one has
\begin{align*}
(*)
=&\psi^i_l(y)-\sum_{k=0}^{l-1}\coeffbin{l}{k}a_{l-k}
\Dy^{l-k}\psi^i_{k}(y)+\theta^i_l+\\
&+\sum_{j=0}^{l-1}\gamma_{l,j}
\Big(
\psi^i_j(y)-\sum_{k=0}^{j-1}\coeffbin{j}{k}a_{j-k}
\Dy^{j-k}\psi^i_{k}(y)+\theta^i_j\Big)+A\phi^i_{l,M}(y)\\
=&
\psi^i_l(y)-\sum_{k=0}^{l-1}\coeffbin{l}{k}a_{l-k}
\Dy^{l-k}\psi^i_{k}(y)
+\sum_{j=0}^{l-1}\gamma_{l,j}
\psi^i_j(y)\\
&-\sum_{j=0}^{l-1}\gamma_{l,j}
\sum_{k=0}^{j-1}\coeffbin{j}{k}a_{j-k}
\Dy^{j-k}\psi^i_{k}(y)
+\theta^i_l+\sum_{j=0}^{l-1}\gamma_{l,j}\theta^i_j+A\phi^i_{l,M}(y)\\
=&\psi^i_l(y)+\sum_{k=0}^{l-1}\Big(-
\coeffbin{l}{k}a_{l-k}\Dy^{l-k}+\gamma_{l,k}
-\sum_{j=k+1}^{l-1}\gamma_{l,j}\coeffbin{j}{k}a_{j-k}
\Dy^{j-k}
\Big)
\psi^i_k(y)+\\
 &+\theta^i_l+\sum_{j=0}^{l-1}\gamma_{l,j}\theta^i_j+A\phi^i_{l,M}(y).
\end{align*}
By the definition of the $\gamma_{j,k}$'s and $\phi^i_{l,M}(y)$'s, each coefficients in the above (first) sum is null and the sum of the last three terms is zero, so that $(*)=\psi^i_l(y)$. Let us discuss \eqref{est-T}. By using \eqref{stima-B} and the fact that $|y_{\pm M}-y_i|\leq 2M\Dy$, since $\Dy(1+2M)\geq 1$ we can write
$$
|T^{\pm M}_{l,i}(y)|\leq
2(\beta\pm\alpha)\Dy^l(1+2M)^{l}\Big(1+\sum_{j=0}^{l-1}|\gamma_{l,j}|\Big).
$$
It remains to prove that, under our constraints, $\sum_{j=0}^{l-1}|\gamma_{l,j}|\leq 1$.  For every $k=0,1,\ldots,l-1$ and $j=k+1,\ldots,l-1$ we consider the estimates
$$
\coeffbin{l}{k}\leq 2^l\quad\mbox{and}\quad \coeffbin{j}{k}\leq 2^l.
$$
By inserting in \eqref{gamma-lk} we can write
$$
|\gamma_{l,k}|\leq 2^l
|a_{l-k}|\Dy^{l-k}+
2^l\sum_{j=k+1}^{l-1}|\gamma_{l,j}| \,|a_{j-k}|
\Dy^{j-k},\quad
k=l-1,\ldots,0.
$$
We also notice that, for $m\geq 1$ and $\Dy\leq 1$,
$$
|a_m|\Dy^m\leq 2(\beta\Dy^2+|\alpha|\Dy)
$$
so that we get
$$
|\gamma_{l,k}|\leq 2^{l+1}
(\beta\Dy^2+|\alpha|\Dy)+
2^{l+1}(\beta\Dy^2+|\alpha|\Dy)\sum_{j=k+1}^{l-1}|\gamma_{l,j}|, \quad
k=l-1,\ldots,0.
$$
Now, if $l2^{l+2}(\beta\Dy^2+|\alpha|\Dy)\leq 1$ we get
$$
|\gamma_{l,k}|\leq \frac{1}{2l}
+\frac{1}{2l}
\sum_{j=k+1}^{l-1}|\gamma_{l,j}|, \quad
k=l-1,\ldots,0
$$
and by summing over $k$,
$$
\sum_{k=0}^{l-1}|\gamma_{l,k}|
\leq \frac{1}{2}
+\frac{1}{2l}
\sum_{k=0}^{l-1}\sum_{j=k+1}^{l-1}|\gamma_{l,j}|
= \frac{1}{2}
+\frac{1}{2l}
\sum_{j=1}^{l-1}\sum_{k=0}^{j-1}|\gamma_{l,j}|
\leq \frac{1}{2}
+\frac{1}{2}
\sum_{j=0}^{l-1}|\gamma_{l,j}|
$$
from which it follows that $\sum_{k=0}^{l-1}|\gamma_{l,k}|\leq 1$ and the statement holds.
\cvd

\begin{remark}\label{A-psi-nostri}
For further use, we write down explicitly the vector $A^{-1}\psi_l^i(y)$ for $l=1,2,4$. Straightforward computations give the following:
\begin{align*}
A^{-1}\psi_1^i(y)
=&\psi^i_1(y)+2\alpha\Dy\Vunita + \phi^i_{1,M}(y),\\
A^{-1}\psi_2^i(y)
=&\psi^i_2(y)+4\alpha\Dy\psi^i_1(y)+2(\beta+2\alpha)\Dy^2\Vunita+\phi^i_{2,M}(y),\\
A^{-1}\psi_4^i(y)
=&
\psi_4^i(y)
+8\alpha\Dy
\psi_3^i(y)
+12(\beta+4\alpha^2)\Dy^2
\psi_2^i(y)
+8(\alpha+12\alpha^2+18\alpha\beta)\Dy^3
\psi_1^i(y)+\\
&+2(\beta+16\alpha^2+96\alpha^3+12\beta^2+192\alpha^3\beta)\Dy^4
\Vunita + \phi^i_{4,M}(y).
\end{align*}
In particular, since $(\psi_l^i(y))_i=0$ for every $l\geq 1$, the $i$th entry of the above sequences are given by
$$
\begin{array}{rl}
(A^{-1}\psi_1^i(y))_i
&=2\alpha\Dy+(\phi^i_{1,M}(y))_i,\\
(A^{-1}\psi_2^i(y))_i
&=2(\beta+2\alpha)\Dy^2+(\phi^i_{2,M}(y))_i,\\
(A^{-1}\psi_4^i(y))_i
&=
2(\beta+16\alpha^2+96\alpha^3+12\beta^2+192\alpha^3\beta)\Dy^4+(\phi^i_{4,M}(y))_i.
\end{array}
$$
By replacing the $\alpha$ and $\beta$ expressions \eqref{alphabeta}, we get the
formulas
\begin{equation}\label{pippoA}
\begin{array}{rl}
(A^{-1}\psi_1^i(y))_i
=&h\mu_Y(v)+(\phi^i_{1,M})_i,\\
(A^{-1}\psi_2^i(y))_i
=&h\bar\rho^2v+2h\Dy\mu_Y(v)+(\phi^i_{2,M})_i,\\
(A^{-1}\psi_4^i(y))_i
=&
h\Dy^2\bar\rho^2v+8h^2\Dy^2\mu_Y(v)^2+24h^3\mu_Y(v)^3
+6h^2\bar\rho^4v^2+\\
&+24\frac{h^4}{\Dy}\,\bar\rho^2v\mu_Y(v)^3+(\phi^i_{4,M})_i.
\end{array}
\end{equation}
\end{remark}

Furthermore, to deal with the numerical boundary conditions, as those given in \eqref{num_boundary_cond_impl}, we need to study the behavior of the $i$th component of the boundary term $\phi^i_{l,M}(y)$ in \eqref{boundary_term} as $i$ is ``far from the boundary'' and $l=1,2,4$.
Here, we use a quite general result (allowing one to set up different boundary conditions) whose precise statement and proof are postponed in Appendix \ref{app-boundary}.

\begin{proposition}\label{prop-boundary}
Suppose that $\beta>|\alpha|$. Let $l\in\N$, $i\in\mathcal{J}_M$ and let $\phi^i_{l,M}(y)$ denote the boundary term in \eqref{boundary_term}. Assume that
$\Dy\leq 1$, $M\Dy\geq 1$ and $l2^{l+2}(\beta\Dy^2+|\alpha|\Dy)\leq 1$. Then one has
$$
|(\phi^i_{l,M}(y))_i|\leq 8(\beta+|\alpha|)\Dy^l(1+2M)^l
\Big(1-\frac{1}{1+\beta+|\alpha|}\Big)^{M-i}.
$$
\end{proposition}
\textbf{Proof.}
Since $\beta>|\alpha|$, $A$ satisfies the requirements in Proposition \ref{prop_boundary_stima}. So, we use such result, that has been specialized to our type of matrix in Remark \ref{bordo-A}:  taking $a=1+2\beta$, $b=-\beta+\alpha$ and $c=-\beta-\alpha$, we obtain
$$
|(A^{-1}\e_{\pm M})_i|\leq \Big(\frac{\beta\pm \alpha}{\gamma^*_{\pm M}}\Big)^{M-i}
$$
where
$$
\gamma^*_{\pm M}=\min\Big(1+2\beta-\frac{2\beta(\beta\mp \alpha)}{1+2\beta},
\frac{1+2\beta+\sqrt{(1+2\beta)^2-4(\beta^2-\alpha^2)}}2\Big).
$$
Straightforward computations give $\gamma^*_{\pm M}\geq 1+\beta\pm \alpha$, so that
$$
|(A^{-1}\e_{\pm M})_i|\leq \Big(\frac{\beta\pm \alpha}{1+\beta\pm\alpha}\Big)^{M-i}.
$$
Now, since $\beta>|\alpha|$ we can write $\frac{\beta\pm \alpha}{1+\beta\pm\alpha}=1-\frac{1}{1+\beta\pm\alpha}\leq
1-\frac{1}{1+\beta+|\alpha|}$, so that
$$
|(A^{-1}\e_{\pm M})_i|\leq \Big(1-\frac{1}{1+\beta+|\alpha|}\Big)^{M-i}.
$$
We insert now the above estimate and estimate \eqref{est-T} in \eqref{boundary_term}, and we get the result.
\cvd



\subsubsection{The case $v<\epsilon$}\label{sect-explicit}

Here we go through our procedure for the explicit in time approximation. We recall that, for $v<\epsilon$, we consider the forward finite differences \eqref{discr_eq_expl_p}
or the backward finite differences \eqref{discr_eq_expl_m} for the first order term depending on the sign of the reaction coefficient: $\mu_Y(v)>0$ or $\mu_Y(v)<0$ respectively, and from \eqref{alphabeta} this means that $\alpha>0$ or $\alpha<0$ respectively.
So, by considering also the case $\alpha=0$ and by coupling with the boundary conditions \eqref{num_boundary_cond_expl}, straightforward computations give that the solution $u^{n}$ of the explicit in time scheme satisfies the condition
$$ u^{n}=C u^{n+1} $$
where
\begin{equation}\label{matrixC}
C = \left(
\begin{array}{ccccc}
1-2\beta-2|\alpha| & 2\beta+2|\alpha| & & & \\
\beta+2|\alpha|\I_{\{\alpha<0\}} & 1-2\beta-2|\alpha| & \beta+2|\alpha|\I_{\{\alpha>0\}} & & \\
 & \ddots & \ddots & \ddots &  \\
 &   & \beta+2|\alpha|\I_{\{\alpha<0\}} & 1-2\beta-2|\alpha| & \beta+2|\alpha|\I_{\{\alpha>0\}} \\
 &   &        &  2\beta+2|\alpha|  & 1-2\beta-2|\alpha|
\end{array}
\right),
\end{equation}
$\alpha$ and $\beta$ being given in \eqref{alphabeta} and $\I$ denoting the indicator function ($\I_{\Gamma}=1$ on $\Gamma$ and $\I_{\Gamma}=0$ otherwise). We remark that $C$ is a stochastic matrix if and only if
\begin{equation}\label{condC}
2\beta+2|\alpha| \leq 1.
\end{equation}
We also notice that if $\epsilon$ is small enough such a condition is not restrictive, but
we will discuss deeper this point later.

In next Section \ref{sect-convMC} we need to deal with $C\psi^i_l(y)$ for $l\leq 4$ and $i\in\mathcal{J}_M$, where the function
$\psi^i_l(y)\in\R^{2M+1}$ are defined in \eqref{psi}. So, we get

\begin{lemma}\label{Cpsi}
Let $\psi^i_l(y)$ be defined in \eqref{psi}. Then for every $l\in\N$ and $i\in\mathcal{J}_M$ one has
$$
C\psi^i_l(y)=\psi^i_l(y)+\sum_{j=0}^{l-1}\coeffbin{l}{j}d_{l-j}\Dy^{l-j}
\psi^i_j(y)+c^{-M}_{l,i}
\e_{-M}+c^M_{l,i}\e_M,
$$
where
$$
d_n=(-1)^n(\beta+2|\alpha|\I_{\{\alpha<0\}})+\beta+2|\alpha|\I_{\{\alpha>0\}},\quad n\in\N,
$$
that is $d_n=2(\beta+|\alpha|)$ if $n$ is even and $d_n=2\alpha$ if $n$ is odd, and $c^{\pm M}_{l,i}$ are given by
$$
c^{\pm M}_{l,i}
=2(\beta+2|\alpha|\I_{\pm \alpha>0})
\sum_{j=0,\,l-j\in \mathcal{A}_\pm}^{l-1}\coeffbin{l}{j}\Dy^{l-j}(y_{\pm M}-y_i)^j
$$
where $\mathcal{A}_+$ and $\mathcal{A}_-$  denotes the set of the even and the odd numbers respectively.
\end{lemma}

The proof is straightforward (it suffices to follow the same arguments developed for Lemma \ref{Apsi}), so we omit it.

\begin{remark}
In the special case $l=1,2,4$, Lemma \ref{Cpsi} gives
$$
\begin{array}{rl}
C\psi_1^i(y)
=& \psi^i_1(y)+2\alpha\Dy\Vunita+c^{-M}_{1,i}\e_{-M}+c^M_{1,i}\e_M,\\
C\psi_2^i(y)
=& \psi^i_2(y)+4\alpha\Dy\psi^i_1(y)+2(\beta+|\alpha|)\Dy^2\Vunita+c^{-M}_{2,i}\e_{-M}+c^M_{2,i}\e_M\\
C\psi_4^i(y)
=&\psi^i_4(y)+8\alpha\Dy\psi^i_3(y)+12(\beta+|\alpha|)\Dy^2\psi^i_2(y)+8\alpha\Dy^3\psi^i_1(y)
+2(\beta+|\alpha|)\Dy^4\Vunita+\\
& +c^{-M}_{4,i}\e_{-M}+c^M_{4,i}\e_M.
\end{array}
$$
In particular, since $(\psi_l^i(y))_i=0$ for every $l\geq 1$ and assuming $|i|<M$, the $i$th entries of the above sequences are given by
\begin{equation}\label{pippoC}
\begin{array}{rl}
(C\psi_1^i(y))_i
&=h\mu_Y(v),\\
(C\psi_2^i(y))_i
&=h\bar\rho^2 v+ h\Dy|\mu_Y(v)|,\\
(C\psi_4^i(y))_i
&=h\Dy^2\bar\rho^2 v + h\Dy^3|\mu_Y(v)|,
\end{array}
\end{equation}
in which we have inserted the formulas for $\alpha$ and $\beta$ in \eqref{alphabeta},
\end{remark}
\subsection{The associated Markov chain and the convergence of the hybrid algorithm}\label{sect-convMC}

We denote, as in Section \ref{sect-tree-fd}, $(\bar V^h_n)_{n=0,1,\ldots,N}$, with $Nh=T$, the  Markov chain approximating the volatility process $V$ introduced in \cite{bib:acz}. We recall that the state-space at step $n$ is given by $\mathcal{V}^h_n$ defined in \eqref{state-space-V}.
We define now the $Y$-component of our Markov chain.

First, given the time-step $h$, we set up the dependence on $h$ for the space-step $\Dy$, the number $M$ giving the points of the grid $\mathcal{Y}_M=\{y_i=Y_0+i\Dy\,;\,i=-M,\ldots,M\}$ and the threshold $\epsilon$ that allows us to use the explicit or the implicit finite difference method. So,  we assume that
\begin{equation}\label{dep-h}
\Dy\equiv \Dy_h=c_yh^p,\qquad M\equiv M_h=c_Mh^{-q},\qquad \epsilon \equiv \epsilon_h=c_\epsilon h^p
\end{equation}
where $c_M>0$ and the constants $c_y,c_\epsilon, p,q>0$ are chosen as follows
\begin{equation}\label{dep-par}
\begin{aligned}
&p<1,\qquad q>p,\qquad \frac{2c_y}{\bar\rho^2}\,\big|r-\delta-\frac{\rho}{\sigma}\kappa\theta\big|<c_\epsilon,\quad\mbox{or}\\
&p=1,\qquad q>p,\qquad \frac{2c_y}{\bar\rho^2}\,\big|r-\delta-\frac{\rho}{\sigma}\kappa\theta\big|<c_\epsilon
<\Big(\frac12-\frac{1}{c_y}\big|r-\delta-\frac{\rho}{\sigma}\kappa\theta\big|\Big)\frac{c_y^2}{\bar\rho^2},
\end{aligned}
\end{equation}
in which the parameters $\kappa,\theta,r,\delta,\sigma$ and $\rho$ come from our model, see Section \ref{sect-model}.
Let us stress that the last constraint in \eqref{dep-par} can be really satisfied, for example by choosing $c_y>4|r-\delta-\frac{\rho}{\sigma}\kappa\theta|$.
We also notice that \eqref{dep-h} and \eqref{dep-par} give  $M_h\Dy_h=O(h^{-q+p})\to\infty$ as $h\to 0$, so that $\mathcal{Y}^h_{M_h}=\{y_i=Y_0+i\Dy_h\,;\,i=-M_h,\ldots,M_h\}\uparrow \R$ as $h\to 0$. Moreover, one has

\begin{proposition}\label{ok}
Let \eqref{dep-h} and \eqref{dep-par} hold and let $\beta=\beta_h$ and $\alpha=\alpha_h$ be given in \eqref{alphabeta} with the constraints \eqref{dep-h} and \eqref{dep-par}. Then there exists $h_0>0$ such that for every $h<h_0$ one has
\begin{itemize}
\item[$i)$] if $v>\epsilon_h$  then $\beta_h>|\alpha_h|$;
\item[$ii)$] if $v\leq \epsilon_h$  then $2\beta_h+2|\alpha_h|<1$.
\end{itemize}

\end{proposition}

\textbf{Proof.} By formula \eqref{drifts}, we write $|\mu_Y(v)|\leq a_Y+b_Yv$, with $a_Y=
|r-\delta-\frac{\rho}{\sigma}\kappa\theta|$ and $b_Y=|\frac{\rho}{\sigma}\kappa-\frac 12|$.

\smallskip

$i)$ One has $\beta_h=\frac {h^{1-2p}}{2c_y^2} \bar\rho^2 v$ and $|\alpha_h|\leq \frac{h^{1-p}}{2c_y}(a_Y+b_Yv)$, so $\beta_h>|\alpha_h|$ if $\frac {h^{1-2p}}{2c_y^2} \bar\rho^2 v> \frac{h^{1-p}}{2c_y}(a_Y+b_Y v)$, that is if
$v(\frac{\bar\rho^2}{c_y}-h^pb_Y)>a_Yh^{p}$. Since $v>\epsilon_h=c_\epsilon h^p$, we get
that $\beta_h>|\alpha_h|$ if $c_\epsilon(\frac{\bar\rho^2}{c_y}-h^pb_Y)>a_Y$, and this holds for every $h$ small because, by  \eqref{dep-par}, one has $c_\epsilon\frac{\bar\rho^2}{c_y}>2a_Y$.

\smallskip

$ii)$ One has $2\beta_h+2\alpha_h\leq \frac{h^{1-2p}}{c_y^2}\bar\rho^2c_\epsilon h^p+\frac{h^{1-p}}{c_y}(a_Y+b_Yc_\epsilon h^p)
=h^{1-p}(\frac{\bar\rho^2}{c_y^2}c_\epsilon +\frac{a_Y}{c_y}+\frac{b_Y}{c_\epsilon}h^p)$. If $p<1$, $2\beta_h+2\alpha_h<1$ for every $h$ small. If instead $p=1$, one has
$2\beta_h+2\alpha_h\leq \frac{\bar\rho^2}{c_y^2}c_\epsilon +\frac{a_Y}{c_y}+\frac{b_Y}{c_\epsilon}h$ and if \eqref{dep-par} holds then
$2\beta_h+2\alpha_h<\frac 12+\frac{b_Y\bar\rho^2}{2a_Yc_y}h<1$ for every $h$ small enough.
\cvd

Now, Proposition \ref{ok} ensures that there exists  $h_0>0$ such that for every $h<h_0$  and for every $v\in \cup_{n=0}^N\mathcal{V}^h_n$, the matrix $A^{-1}$ discussed in Section \ref{sect-implicit}
and the matrix $C$ discussed in Section \ref{sect-explicit} are both well defined and are stochastic matrices. So, for $h$ small, that is $h<h_0$ with $h_0$ as in Proposition \ref{ok}, we define $\Pi^h(v)$ as follows:
\begin{itemize}
\item[-]
if $v>\epsilon_h$, $\Pi^h(v)$ is the inverse of the matrix $A$ in \eqref{matrixA},
\item[-]
if $v\leq \epsilon_h$, $\Pi^h(v)$ is the matrix $C$ in \eqref{matrixC}.
\end{itemize}

As a consequence, we can assert that for every $v\in \cup_{n=0}^N\mathcal{V}^h_n$, $\Pi^h(v)=(\Pi^h(v)_{ij})_{i,j\in\mathcal{J}_{M_h}}$ is a stochastic matrix.
We now define the following transition probability law: at time-step $n\in\{0,1,\ldots,N\}$, for $(y_i,v_{n,k})\in\mathcal{Y}^h_{M_h}\times\mathcal{V}_n^h$ we set $\mu^h(y_i,v_{n,k};\cdot)$ the probability law in $\R^2$ as in \eqref{mu-h}, that is
\begin{equation}\label{mu-h-bis}
\mu^h(y_i,v_{n,k};A\times B)
=\sum_{k^*\in\{k_u^h(n,k),k_d^h(n,k)\}}\sum_{j\in\mathcal{J}_{M_h}}\Pi^h(v_{n,k})_{i,j}p^h_{k^*}\delta_{\{y_j\}}(A)\delta_{\{v_{n+1,k^*}\}}(B).
\end{equation}
So, we call $\bar X^h=(\bar X^h_n)_{n=0,1,\ldots,N}$ the $2$-dimensional Markov chain having \eqref{mu-h-bis} as its transition probability law at time-step $n\in\{0,1,\ldots,N\}$, that is
$$
\P\big(\bar X^h_{n+1}=(y_j,v_{n+1,k^*})\mid \bar X^h_{n}=(y_i,v_{n,k})\big)
=\begin{cases}
\Pi(v_{n,k})_{ij}\,p^h_{k^h_u(n,k)} &\quad\mbox{if } k^*=k^h_u(n,k)\\
\Pi(v_{n,k})_{ij}\,p^h_{k^h_d(n,k)} &\quad\mbox{if } k^*=k^h_d(n,k)\\
0 &\quad\mbox{otherwise},
\end{cases}
$$
for every $(y_i,v_{n,k})\in\mathcal{Y}^h\times \mathcal{V}_n^h$ and $(y_j,v_{n+1,k^*})\in\mathcal{Y}^h\times \mathcal{V}_{n+1}^h$.
Since $\sum_{j}\Pi(v)_{ij}=1$, one gets that the second component of $\bar X^h$ is a Markov chain itself and it equals, in law, to $\bar V^h$. So, we write $\bar X^h=(\bar Y^h,\bar V^h)$ and for every function $f\,:\,\R^2\to\R$ we have
\begin{equation}\label{media-h}
\E(f(\bar Y^h_{n+1}, \bar V^h_{n+1})\mid
(\bar Y^h_{n},\bar V^h_n)=(y_i,v_{n,k})\big)
=\sum_{k^*\in\{k_u^h(n,k),k_d^h(n,k)\}}\!\sum_{j\in\mathcal{J}_{M_h}}
f(y_j,v_{n+1,k^*})\Pi^h(v_{n,k})_{i,j}p^h_{k^*}.
\end{equation}

Coming back to the discussion in Section \ref{sect-tree-fd}, by \eqref{media-h} we can assert that our algorithm actually approximates the diffusion pair $X=(Y,V)$ with the Markov chain $\bar X^h=(\bar Y^h,\bar V^h)$. So, we set $X^h=(Y^h,V^h)$ as the piecewise constant and c\`adl\`ag interpolation in time of $\bar X^h$, that is
\begin{equation}\label{Xh}
X^h_t=\bar X^h_n,\quad t\in[nh,(n+1)h),\quad
n=0,1,\ldots,N-1.
\end{equation}
We set $\mathrm{D}([0,T];\R^2)$ the space of the $\R^2$-valued and c\`adl\`ag functions on the interval $[0,T]$, that  we assume to be endowed with the Skorohod topology (see e.g. Billingsley \cite{bill}). Our main result is the following:

\begin{theorem}\label{th-conv}
Under \eqref{dep-h} and \eqref{dep-par}, as $h\to 0$ the family $\{X^h\}_h=\{(Y^h,V^h)\}_h$ defined through \eqref{Xh} and \eqref{media-h} weakly converges in the space $\mathrm{D}([0,T];\R^2)$ to the diffusion process $X=(Y,V)$ solution to \eqref{Yt1}-\eqref{vt1}.
\end{theorem}

\textbf{Proof.}
The idea of the proof is standard, see e.g. Nelson and Ramaswamy \cite{nr} or also classical books such as Billingsley \cite{bill}, Ethier and Kurtz \cite{ek} or Stroock and Varadhan \cite{sw}.

\smallskip

\def\dT{h}
To simplify the notations, let us set
\begin{align*}
&\M^Y_{n,i,k}(h;l)
=\E\big((\bar Y^\dT_{n+1}-y_i)^l\,|\,(\bar Y^h_{n},\bar V^h_n)=(y_i,v_{n,k})\big), \quad l=1,2,4,\\
&\M^V_{n,i,k}(h;l)=\E\big((\bar V^\dT_{n+1}-v_{n,k})^l\,|\,(\bar Y^h_{n},\bar V^h_n)=(y_i,v_{n,k})\big), \quad l=1,2,4,\\
&\M^{Y,V}_{n,i,k}(h)= \E\big((\bar Y^\dT_{n+1}-y_i)(\bar V^\dT_{n+1}-v_{n,k})\,|\,(\bar  Y^h_{n},\bar  V^h_n)=(y_i,v_{n,k})\big).
\end{align*}
It is clear that $\M^Y_{n,i,k}(h;l)$ is the local moment of order $l$ at time $n\dT$ related to $Y$, as well as $\M^{V}_{n,k}(h;l)$ is related to the component $V$, and $\M^{Y,V}_{n,i,k}(h)$ is the local cross-moment of the two components at the generic time step $n$. So, the proof reduces in checking that, for fixed $r_*, v_*>0$ and setting $\Lambda_*=\{(n,i,k)\,:\,v_{n,k}\leq v_*,|y_{i}|\leq r_*\}$, then the following properties $i)$, $ii)$ and $iii)$ hold:

\medskip

$i)$ (convergence of the local drift)
\begin{align*}
&\lim_{\dT\to 0} \sup_{(n,i,k)\in \Lambda_*}\frac 1\dT\,\big|\M^Y_{n,i,k}(h;1)-(\mu_Y)_{n,k}\dT\big|=0,\\
&\lim_{\dT\to 0} \sup_{(n,i,k)\in \Lambda_*}\frac 1\dT\,\big|\M^V_{n,i,k}(h;1)-(\mu_V)_{n,k}\dT|=0;
\end{align*}
where we have set $(\mu_Y)_{n,k}=\mu_Y(v_{n,k})$ and $(\mu_V)_{n,k}=\mu_V(v_{n,k})$;

\medskip

$ii)$ (convergence of the local diffusion coefficient)
\begin{align*}
&\lim_{\dT\to 0} \sup_{(n,i,k)\in \Lambda_*}\frac 1\dT\,\big|\M^Y_{n,i,k}(h;2)-\bar\rho^2v_{n,k}\dT\big|=0,\\
&\lim_{\dT\to 0} \sup_{(n,i,k)\in \Lambda_*}\frac 1\dT\,\big|\M^V_{n,k}(h;2)-\sigma^2 v_{n,k}\dT\big|=0\\
&\lim_{\dT\to 0} \sup_{(n,i,k)\in \Lambda_*}\frac 1\dT\,\big|\M^{Y,V}_{n,i,k}(h)\big|=0;
\end{align*}

\medskip

$iii)$ (fast convergence to 0 of the fourth order local moments)
\begin{align*}
&\lim_{\dT\to 0} \sup_{(n,i,k)\in \Lambda_*}\frac 1\dT\,\M^Y_{n,i,k}(h;4)=0,\\
&\lim_{\dT\to 0} \sup_{(n,i,k)\in \Lambda_*}\frac 1\dT\,\M^{V}_{n,i,k}(h;4)=0.
\end{align*}

We recall that the $V$-component of the 2-dimensional Markov chain is a Markov chain itself and we have
$$
\M^V_{n,i,k}(h;l)\equiv
\M^V_{n,k}(h;l)
=\E\big((\bar V^\dT_{n+1}-v_{n,k})^l\,|\,\bar V^h_n=v_{n,k}\big), \quad l=1,2,4.
$$
The limits in $i)$, $ii)$ and $iii)$ containing $\M^V_{n,k}(h;l)$ for $l=1,2,4$, have been already proved in \cite{bib:acz} (see Theorem 7 therein), so we prove the validity of the remaining limits.

We set $\psi^i_{l}(y)$ the vector in $\R^{2M_h+1}$ given by $(\psi^i_{l}(y))_j=(y_j-y_i)^l$, $j\in\mathcal{J}_{M_h}$. From \eqref{media-h} we get
$$
\M^Y_{n,i,k}(h;l)
=\sum_{y_j\in\mathcal{Y}^h}\Pi^h(v_{n,k})_{i,j}(\psi^i_{l}(y))_j
=\big(\Pi^h(v_{n,k})\psi^i_{l}(y)\big)_{i}
$$
and we notice that the above quantity has been already discussed in the previous sections.
We set $\Lambda_*=\Lambda_{*,1,h}\cup \Lambda_{*,2,h}$, with
\begin{align*}
&\Lambda_{*,1,h}=\{(n,i,k)\,:\,\epsilon_h<v_{n,k}\leq v_*,|y_{i}|\leq r_*\},\\
&\Lambda_{*,2,h}=\{(n,i,k)\,:\,v_{n,k}\leq \epsilon_h,|y_{i}|\leq r_*\}.
\end{align*}
For $(n,i,k)\in \Lambda_{*,1,h}$, $\Pi^h(v_{n,k})$ is the inverse of the matrix $A$ in \eqref{matrixA}. So, by using \eqref{pippoA}, we have
$$
\begin{array}{rl}
\M^Y_{n,i,k}(h;1)
=(A^{-1}\psi_1^i(y))_i
=&h(\mu_Y)_{n,k}+(\phi^i_{1,M_h})_i,\\
\M^Y_{n,i,k}(h;2)
=(A^{-1}\psi_2^i(y))_i
=&h\bar\rho^2v_{n,k}+2h\Dy_h(\mu_Y)_{n,k}+(\phi^i_{2,M_h})_i,\\
\M^Y_{n,i,k}(h;4)
=(A^{-1}\psi_4^i(y))_i
=&
h\Dy_h^2\bar\rho^2v_{n,k}+8h^2\Dy_h^2(\mu_Y)_{n,k}^2+24h^3(\mu_Y)_{n,k}^3
+6h^2\bar\rho^4v_{n,k}^2+\\
&+24\frac{h^4}{\Dy_h}\,\bar\rho^2v_{n,k}(\mu_Y)_{n,k}^3+(\phi^i_{4,M_h})_i,
\end{array}
$$
$\phi^i_{l,M_h}(y)$ being given in \eqref{boundary_term}.
In Lemma \ref{lemma_stima_bound} below, we prove that for $l\leq 4$,
$$
\sup_{(n,i,k)\in \Lambda_{*,1,h}}\frac 1h\big|(\phi^i_{l,M_h}(y))_i|\to 0 \quad\mbox{as}\quad h\to 0.
$$
And since $(\mu_Y)_{n,k}$ is bounded on $\Lambda_{*}$, the limits in $i)$, $ii)$ and $iii)$ associated to $\M^Y_{n,i,k}(h;l)$, $l=1,2,4$, hold uniformly in $\Lambda_{*,1,h}$. We prove the same on the set $\Lambda_{*,2,h}$. For $(n,i,k)\in \Lambda_{*,2,h}$, the matrix $\Pi^h(v_{n,k})$ to be taken into account is given by the matrix $C$ in \eqref{matrixC}. Moreover, for $h$ small enough, we have that if $(n,i,k)\in \Lambda_{*,2,h}$ then $|i|<M_h$. Therefore, by \eqref{pippoC} we obtain
$$
\begin{array}{rl}
\M^Y_{n,i,k}(h;1)
=(C\psi_1^i(y))_i
&=h(\mu_Y)_{n,k},\\
\M^Y_{n,i,k}(h;2)
=(C\psi_2^i(y))_i
&=h\bar \rho^2 v_{n,k}+ h\Dy_h|(\mu_Y)_{n,k}|,\\
\M^Y_{n,i,k}(h;4)
=(C\psi_4^i(y))_i
&=h\Dy_h^2\bar \rho^2 v_{n,k} + h\Dy_h^3|(\mu_Y)_{n,k}|
\end{array}
$$
and again the limits in $i)$, $ii)$ and $iii)$ concerning $\M^Y_{n,i,k}(h;l)$, $l=1,2,4$, hold uniformly in $\Lambda_{*,2,h}$. It remains to study the cross-moment. By using \eqref{media-h}, it is given by
$$
\M^{Y,V}_{n,i,k}(h)=\M^Y_{n,i,k}(h;1)\M^{V}_{n,k}(h;1)
$$
and the convergence as in $ii)$ immediately follows from the already proved limits in $i)$.
\cvd

In order to conclude, we only need to prove next
\begin{lemma}\label{lemma_stima_bound}
Assume that \eqref{dep-h} and \eqref{dep-par} both hold. Let $v_*,r_*>0$ and set
$$
\Lambda_{*,1,h}=\Lambda_*=\{(n,i,k)\,:\,\epsilon_h<v_{n,k}\leq v_*,|y_{i}|\leq r_*\}
$$
Then one has
\begin{equation}\label{stima_boundary}
\lim_{h\to 0}\frac 1h\sup_{(n,i,k)\in \Lambda_{*,1,h}}\Frac{1}{h}|(\phi^i_{l,M_h}(y))_i|
=0,
\end{equation}
for every $l\leq 4$, where $\phi^i_{l,M_h}(y)$ is defined in \eqref{boundary_term}
with $M=M_h$.
\end{lemma}

\textbf{Proof.} We use Proposition \ref{prop-boundary}. Here, $C$ denotes a positive constant that may vary from line to line.

Under \eqref{dep-h} and \eqref{dep-par}, for $(n,i,k)\in \Lambda_{*,1,h}$ we have already observed that $\beta_h>|\alpha_h|$, $\alpha_h,\beta_h$ being given in \eqref{alphabeta}, and the constraints $\Dy_h\leq 1$ and $M_h\Dy_h\geq 1$ are trivially satisfied for $h$ small. Moreover, on the set $\Lambda_{*,1,h}$, there exists $C>0$ such that
for every $l\leq 4$,
$$
l2^{l+2}(\beta_h\Dy_h^2+|\alpha_h|\Dy_h)
\leq C h\leq 1.
$$
for $h$ small enough. And since $\beta_h+|\alpha_h|\leq C h^{1-2p}$, we also have
$$
1-\frac 1{1+\beta+|\alpha|}\leq 1-\frac 1{1+C h^{1-2p}}.
$$
Then, by applying Proposition \ref{prop-boundary}, we can write
$$
|(\phi^i_{l,M}(y))_i|\leq
Ch^{-(q-p)l+1-2p}\Big(1-\frac 1{1+C h^{1-2p}}\Big)^{M_h-i}.
$$
Now, on the set $\Lambda_{*,1,h}$ we have $|y_i|\leq r_*$, so that $|i|\leq \frac{r_*+|Y_0|}{\Delta y_h}\leq Ch^{-p}$. And by recalling that $q>p$, we get $M_h-i\geq C(h^{-q}-h^{-p})\geq ch^{-q}$ for some $c>0$. So, we have proved  that there exist $C,c>0$  such that for every $h$ close to $0$
$$
\sup_{(n,i,k)\in \Lambda_{*,1,h}}\frac 1h\,|(\phi^i_{l,M}(y))_i|\leq
Ch^{-(q-p)l-2p}\Big(1-\frac 1{1+C h^{1-2p}}\Big)^{ch^{-q}}=:CI_h\quad \mbox{for}\quad l\leq 4.
$$
We see now that, $I_h\to 0$ as $h\to 0$ under our constraints $q>p>0$ and $p\leq 1$. In fact, if $1-2p\geq 0$ then $1-\frac 1{1+C h^{1-2p}}\leq 1-\frac 1{1+C}$ for every $h$ small enough, and $I_h\to 0$. Otherwise, if $1-2p<0$ then we write
\begin{align*}
\log I_h
&\simeq -((q-p)l+2p)\log h-\frac {ch^{-q}}{1+C h^{1-2p}}\\
&=h^{-(q-2p+1)}\Big[-((q-p)l+2p)h^{q-2p+1}\log h-\frac {c}{h^{2p-1}+C}\Big].
\end{align*}
Since $q-2p+1=q-p-p+1>-p+1\geq 0$, then $q-2p+1>0$ and $\log I_h\to -\infty$. The statement now holds.
\cvd

\begin{remark}\label{boundary-cond}
Theorem \ref{th-conv} proves the convergence of the algorithm in the case we introduce suitable boundary conditions  in the finite difference component of our procedure.
We stress that our Neumann-type conditions  \eqref{num_boundary_cond_impl} and \eqref{num_boundary_cond_expl} have brought to the matrices $A$ and $C$ given by \eqref{matrixA} and \eqref{matrixC} respectively. But these conditions
may be replaced by other types of boundary conditions, that can be handled by using Proposition \ref{prop_boundary_stima} (see Appendix \ref{app-boundary}). For example, in the case of time-independent Dirichlet-type conditions,
that is conditions that fix the value of the function at the end-points $y_{\pm M}$ of the grid (this typically appears in financial problems), one could set $u^n_{\pm M}=u^{n+1}_{\pm M}$ for every $n$.
 This means that the matrices $A$ and $C$ to be taken into account become
\begin{equation}\label{matrixA-dirichlet}
A = \left(
\begin{array}{ccccc}
1 & 0 & & & \\
\alpha-\beta & 1+2\beta & -\alpha-\beta & & \\
 & \ddots & \ddots & \ddots &  \\
 &   & \alpha-\beta & 1+2\beta & -\alpha-\beta \\
 &   &        &   0 & 1
\end{array}
\right).
\end{equation}
and
\begin{equation}\label{matrixC-dirichlet}
C = \left(
\begin{array}{ccccc}
1 & 0 & & & \\
\beta+2|\alpha|\I_{\{\alpha<0\}} & 1-2\beta-2|\alpha| & \beta+2|\alpha|\I_{\{\alpha>0\}} & & \\
 & \ddots & \ddots & \ddots &  \\
 &   & \beta+2|\alpha|\I_{\{\alpha<0\}} & 1-2\beta-2|\alpha| & \beta+2|\alpha|\I_{\{\alpha>0\}} \\
 &   &        &  0  & 1
\end{array}
\right),
\end{equation}
respectively. So, again Proposition \ref{prop_boundary_stima} in next Appendix \ref{app-boundary} can be used - for details, see Remark \ref{bordo-A-dirichlet}. As a consequence, the boundary contributions can be handled as in Lemma \ref{lemma_stima_bound}, and the proof of Theorem \ref{th-conv} can be easily reproduced.
\end{remark}

\begin{remark}\label{boundary-cond-bis}
We also recall that the use of a boundary is a  numerical requirement which is necessary to solve the problem in practice. However, one could prove the convergence result even in the whole grid $\mathcal{Y}^h=\{y_i=Y_0+i\Dy_h\}_{i\in\Z}$, that is by considering the solutions to \eqref{discr_eq} and \eqref{discr_eq_expl_p} without linking to these equations any boundary condition. In fact, under the requirements of Theorem \ref{th-conv}, the inverse of the implicit difference linear operator $A$ associated to \eqref{discr_eq} remains a Markovian transition function, as well as the explicit difference linear operator $C$ related to \eqref{discr_eq_expl_p}. Moreover, formulas similar to \eqref{pippoA} and \eqref{pippoC} hold, with a simplification due to the fact that boundary contributions do not exist. For the sake of completeness, we give these proofs in Appendix \ref{app-infinity}. This means that the arguments used to prove Theorem \ref{th-conv} can be applied also on the infinite grid $\mathcal{Y}^h=\{y_i=Y_0+i\Dy_h\}_{i\in\Z}$.
\end{remark}

Thanks to  Theorem \ref{th-conv} we can deal with the convergence of the price evaluated on our Markov chain to the one computed on the Heston model.

The problem of pricing European options is simple when the payoff is not too complicated. In fact, let $T$ denote the maturity date and $f$ be the payoff function, that is $f\,:\,\mathrm{D}([0,T];\R)\to \R_+$. First, we write the payoff in terms of the transformation \eqref{Y}, so we get the transformed payoff-function $g(y,v)=f(e^{y+\frac\rho\sigma v})$, $(y,v)\in \mathrm{D}([0,T];\R^2)$, and the associated option prices on the continuous and the discrete model as seen at time $0$ are given by
$$
\mbox{$P_{Eu}=\E\big(\tilde g(Y,V)\big)\quad$ and
$\quad P^h_{Eu}=\E\big(\tilde g(  Y^h,   V^h)\big)$},
$$
respectively, $\tilde g$ denoting the discounted payoff, i.e. $\tilde g=e^{-rT}g$ (it is  clear that the writing $\E$ for both prices is an abuse of notations, since in principle one should use the notations $\E_\P$ and $\E_{\P^h}$ related to the measures $\P$ and $\P^h$ of the probability space where the processes $(Y,V)$ and $(  Y^h,  V^h)$ are defined). Now, the weak convergence in Theorem \ref{th-conv} ensures the convergence $P^h_{Eu}\to P_{Eu}$ of the European price when the discounted payoff-function fulfills the following requests: $(y,v)\mapsto \tilde g(y,v)$ is continuous and  there exists $a>0$ and $h_*>0$ such that
$$
\sup_{h<h_*}\E\big(|\tilde g(  Y^h,  V^h)|^{1+a}\big)<\infty.
$$
This a consequence of standard results on the convergence of the expectations  for sequences of random variables which are weakly convergent and satisfy uniform integrability properties.

As for American style options, even for simple payoffs things are more difficult because of the presence of optimal stopping times. However, due to the results in Amin and Khanna \cite{ak} (but see also Lamberton and Pag\`es \cite{lp}), we can deduce the convergence of the prices for suitable payoffs. In fact, let  $f\,:\,[0,T]\times \mathrm{D}([0,T];\R) \to \R_+$ denote a payoff function. By passing to the pair $(Y,V)$ and by considering the resulting discounted payoff function $\tilde g(t,y,v)=e^{-rt} f(t,e^{y+\frac\rho\sigma v}) $, the associated option prices on the continuous and the discrete model as seen at time $0$ are given by
$$
\mbox{$P_{Am}=\sup_{\tau\in\mathcal{G}_{0,T}}\E\big(\tilde g(\tau,Y,V)\big)\quad$ and
$\quad P^h_{Am}=\sup_{\eta\in\mathcal{G}_{0,T}^h}\E\big(\tilde g(\eta,  Y^h,   V^h)\big)$},
$$
where $\mathcal{G}_{0,T}$ and $\mathcal{G}_{0,T}^h$ denote the stopping times in $[0,T]$ w.r.t. the filtration $\cl F_t=\sigma((Y_s,V_s)\,:\,s\leq t)$  and $\cl F^h_t=\sigma((  Y_s^h,  V_s^h)\,:\,s\leq t)$ respectively.

Consider the following assumptions:
\begin{itemize}
\item[(H1)]
$\tilde g\,:\,[0,T]\times \mathrm{D}([0,T];\R^2)\to\R_+$ is a continuous function (in the product topology) and for every $\xi,\eta\in \mathrm{D}([0,T];\R^2)$ such that $\xi_s=\eta_s$ for every $s\in[0,t]$ then $\tilde g(t,\xi)=\tilde  g(t,\eta)$;
\item[(H2)]
there exist $a>0$ and $h_*>0$ such that
$\sup_{h<h_*}\E\big(\sup_{t\leq T}|\tilde g(t,  Y^h,  V^h)|^{1+a}\big)<\infty$.
\end{itemize}
Let $ \eta^h$ denote the optimal stopping time related to the discrete problem, that is
$$
\mbox{
$P_{Am}^h=\E\big(\tilde  g( \eta^h,  Y^h,   V^h)\big)$.}
$$
Then by using the arguments in Amin and Khanna \cite{ak},
properties $i)$-$iii)$ in the proof of Theorem \ref{th-conv} ensure the existence of $\eta\in \mathcal{G}_{0,T}$ such that the triple $( \eta^h,  Y^h,  V^h)$ converges in law to  $(\eta, Y, V)$   - this is a very roughly speaking: in the words by Amin and Khanna, $\eta$ is ``in  some appropriate sense a legitimate stopping time w.r.t. the filtration generated by $(Y,V)$'', and to be precise one should refer to a further probability space (for technical details, see the discussion at pages 299-300 and in particular Theorem 4.1 in \cite{ak}). We stress that in \cite{ak} this result is based on two starting assumptions: the (global) Lipschitz continuity and the sublinearity property for both the drift and the diffusion coefficient of the pair $(Y,V)$. Here, the Lipschitz continuity property does not hold because of the presence of the square-root function, so the diffusion coefficient is only H\"older continuous. Nevertheless, they use such a property just to ensure the existence of the strong solution, a condition that here holds, and use specifically the sublinearity property of the drift coefficients, that here holds.  So, their arguments apply as well. Once the weak convergence of the triple is achieved, the plan to prove the convergence of the prices is the following. Under (H1) one gets that $\{\tilde g( \eta^h,   Y ^h,   V^h)\}_h$ converges in law, as $h\to 0$, to $\tilde g(\eta, Y, V)$. Moreover, (H2) implies that the $\{\tilde g( \eta^h,   Y ^h,   V^h)\}_h$ is a uniformly integrable family of random variables, and this suffices to get the convergence of the expectations. Finally, one gets $P_{Am}=\E(\tilde g(\eta,Y,V))$, and the convergence $P^h_{Am}\to P_{Am}$ of the American prices is achieved.

As an immediate consequence, both European and American put options can be numerically evaluated by means of the approximating algorithm  described in Section \ref{sect-tree-fd} (and, due to the call-put parity formula, also European call options can be numerically priced with our method). However, in next Section \ref{sect-numerics} we apply our procedure to numerically price European/American barrier options written on the Heston model, that is options which are much more sophisticated and, for example, do not fulfill (H1). Nevertheless, the numerical experiments give a real evidence of the goodness (that is, convergence) of our algorithm also in this case.

\section{Numerical results}\label{sect-numerics}
In this section we provide numerical results in order to asses the
efficiency and the robustness of our hybrid tree-finite difference method in the case of
plain vanilla options and barrier options. All the computations have
been performed in double precision on a PC 1,7 GHz Intel Core I5 with
4 Gb of RAM.

The hybrid tree-finite difference algorithm introduced in Section \ref{sect-tree-fd} is here called \textbf{HTFD} and it is split in two different variants:
\begin{itemize}
\item
\textbf{HTFD1}: when the number $N_t$ of time steps and the number $N_S$ of space steps are different;
\item
\textbf{HTFD2}: when the number $N_t$ of time steps and the number $N_S$ of space steps are equal.
\end{itemize}

\subsection{European and American vanilla options: comparison with
  tree methods}\label{sect-num-euam}
We compare here the performance of the hybrid tree-finite
difference algorithm \textbf{HTFD} with the tree method of Vellekoop
Nieuwenhuis (called \textbf{VN}) in \cite{vn} for the computation of European and
American options in the Heston model.  All the numerical results of this Section
(except for our method) are obtained by using the Premia software \cite{pr}.
In the European and American option contracts we are dealing with, we
consider the following set of parameters: initial price $S_0=100$, strike price $K=100$,
maturity $T=1$, interest rate $r=\log(1.1)$, dividend rate $\delta=0$, initial volatility $V_0=0.1$, long-mean $\theta=0.1$, speed of mean-reversion   $\kappa=2$, correlation $\rho=-0.5$. In order to study the numerical robustness of the algorithms, we choose
three different values for $\sigma$: we set  $\sigma=0.04, 0.5, 1$.
We first consider the case $\sigma=0.04$, that is $\sigma$ close to zero (which implies that the Heston PDE is convection-dominated in the $V$-direction).  Moreover,
for $\sigma=1$, we stress that the Feller condition $2\kappa \theta\geq\sigma^2$ is not satisfied.

In the (pure) tree method \textbf{VN}, we fix the number of points in the $V$ coordinate as
$N_V=50$, with varying number of time and space steps: $N_t=N_S=50,
100, 200, 400$.

As already mentioned, the numerical study of the hybrid tree-finite difference method \textbf{HTFD} is split in two cases: \textbf{HTFD1} refers to the (fixed) number of time steps $N_t=100$ and varying number of space steps $N_S=50, 100, 200, 400$;
we add the situation \textbf{HTFD2} where the number of  time steps is equal to the
number of space steps $N_t=N_S=50, 100, 200, 400$.

Table \ref{tab1} reports European put option prices. Comparisons are given with a  benchmark value obtained using the Heston closed formula \textbf{CF} in \cite{hes}.

In Table \ref{tab2} we provide results for
American put option prices. In this case we use a  benchmark from the Monte Carlo Longstaff-Schwartz algorithm, called  \textbf{MC-LS}, as in \cite{ls},  with a
huge number of Monte Carlo simulation (1 million iterations) which are done by means of
the accurate Alfonsi \cite{al} discretization scheme for the CIR
process with $M=100$ discretization time steps and bermudan exercise
dates. We recall that the Alfonsi method provides a Monte Carlo weak second-order scheme for the CIR process, without any restriction on its parameters.

Table \ref{tab3} refers to the computational time cost (in seconds) of the different algorithms for $\sigma=0.5$ in the European case.

\begin{table} [ht] \centering
\footnotesize
{\begin{tabular} {@{}cccccc@{}} \toprule & $N_S$ & VN &HTFD1  &HTFD2 & CF \\
\hline
$\sigma=$0.04& 50 &8.040982 &7.934492 &7.911034 &\\
& 100 &8.021780 &7.970437 &7.970437  &\\
& 200 &8.003938  &7.978890 &7.983188 &7.994716\\
& 400 &7.984248  &7.980984 &7.990825 &\\
\hline
$\sigma=$0.50& 50 &8.148234 &7.758954 &7.746533 &\\
& 100 &7.727191  &7.804520   &7.804520 &\\
& 200 &7.813599  &7.816749  &7.821404 &7.8318540\\
& 400 &7.910909  &7.818596 &7.827805 &\\
\hline
$\sigma=$1.00& 50 &6.586889  &7.214303 &7.247748 &\\
& 100 &7.114225  &7.225292 &7.225292 &\\
& 200 &7.964052  &7.228235 &7.229139 &7.2313083\\
& 400 &6.639931  &7.224356 &7.233742 &\\
\end{tabular}}
\caption{\em \small{Prices of European put options. $\sigma=0.04, 0.5,
    1$.  $ S_0=100$,
    $K=100$, $T=1$, $r=\log(1.1)$, $\delta=0$, $V_0=0.1$, $\theta=0.1$, $\kappa=2$, $\rho=-0.5$.}}
\label{tab1}
\end{table}

\begin{table} [ht] \centering
\footnotesize
{\begin{tabular} {@{}cccccc@{}} \toprule & $N_S$ & VN &HTFD1 &HTFD2 & MC-LS \\
\hline
$\sigma=$0.04& 50 &9.100312 &8.966651 &8.932445 &\\
& 100 &9.086233 &9.016732   &9.016732 &\\
& 200 &9.073722  &9.028866 &9.042581 &9.074102\\
& 400 &9.063396  &9.031881 &9.054538 &\\
\hline
$\sigma=$0.50& 50 &9.150887 &8.763369& 8.731867&\\
& 100 &8.892206 &8.841776 &8.841776 &\\
& 200 &8.981855 &8.862606  &8.878530 &8.904514\\
& 400 &9.058313  &8.866911 &8.892583 &\\
\hline
$\sigma=$1.00& 50 &8.588392  &8.185052 &8.206052 &\\
& 100 &9.020989   & 8.263395  &8.263395 &\\
& 200 &9.251595   &8.281755 &8.290371 &8.277985\\
& 400 &9.102788   &8.283214 &8.304415 &\\
\end{tabular}}
\caption{\em \small{Prices of American put options. $\sigma=0.04, 0.5,
    1$.  $ S_0=100$,
    $K=100$, $T=1$, $r=\log(1.1)$, $\delta=0$, $V_0=0.1$, $\theta=0.1$, $\kappa=2$, $\rho=-0.5$.}}
\label{tab2}
\end{table}

\begin{table} [ht] \centering
\footnotesize
{\begin{tabular} {@{}ccccc@{}} \toprule & $N_S$ & VN &HTFD1 &HTDF2  \\
\hline
& 50 & 0.11 &0.02 &0.007 \\
& 100 & 0.42 &0.04 &0.040\\
& 200 & 1.73 &0.08 &0.380\\
& 400 & 7.06 &0.16 &3.040\\
\end{tabular}}
\caption{\em \small{Computational times (in seconds) for European put
    options for $\sigma=0.5.$}}
\label{tab3}
\end{table}

The numerical results show that the hybrid tree-finite difference
method is very accurate, reliable and efficient.

\subsection{American vanilla options: comparison with
  finite difference methods}\label{sect-num-euam-bis}
We compare here the performance of the hybrid tree-finite
difference algorithm \textbf{HTFD} with various finite difference methods pricing results
given in \cite{ClarkeParrott,it,Oosterlee,ForsythVetzalZvan}.
The model parameter values are
 $$
 \kappa=5,\qquad \theta=0.16, \qquad \sigma=0.9,\qquad \rho=0.1,\qquad
 r=0.10, \qquad  \delta=0, \qquad
 V_0=0.25.
 $$
The strike was chosen to be $K=10$ and different values of the
initial stock prices $S_0=8,
9, 10, 11, 12$ are considered.
Comparisons in the American cases are given with the reference values
provided in \textbf{ZFV} (Zven, Forsyth,
Vetzal \cite{ForsythVetzalZvan}), \textbf{IT-PSOR} (Ikonen and Toivanen \cite{it}), \textbf{OO} (Oosterlee \cite{Oosterlee}) and \textbf{CP} (Clarke and Parott \cite{ClarkeParrott}) methods. These values with those obtained by the \textbf{HTFD} algorithm are compared in Table \ref{tab4}.
In order to study the convergence behaviour of  \textbf{HTFD2}, in Table \ref{tab4ratio} we consider the following convergence ratio proposed in \cite{dfl}:
$$
\mathrm{ratio}=\frac{P_{\frac{N}{2}}-P_{\frac{N}{4}}}{P_{N}-P_{\frac{N}{2}}},
$$
where $P_{N}$ is the approximated price obtained with $N=N_t=N_s$ number of  time steps.
Table \ref{tab4ratio} suggests that the convergence ratio for
\textbf{HTDF2} is linear, as it is expected to be because of the tree contribution (whose error typically behaves linearly). Moreover, Table \ref{tab4ratio} shows that the numerically observed ratios are very stable, and this gives a strong evidence of the robustness of the method.

\begin{table} [ht] \centering
\footnotesize
{\begin{tabular} {@{}cccccccc@{}} \toprule & $N_S$ &HTFD1
    &HTFD2  & ZFV &IT-PSOR  &OO &CP  \\
\hline
$S_0 =8$ & 50 &2.077001 &2.076301  & & & &\\
&100 &2.077402 &2.077402  & & & &\\
&200 &2.077510 &2.077904 &2.0784 &2.0783 &2.0790 &2.0733\\
&400 &2.077540&2.078141 & & & &\\
&800 &2.077548&2.078255 & & & &\\
\hline
$S_0 =9$& 50&1.330748 & 1.329763 & & & &\\
&100 &1.332058 &1.332058   & & & &\\
&200 &1.332351 &1.332923  &1.3337 & 1.3335 &1.3340 &1.3290 \\
&400 &1.332424 &1.333295  & & & &\\
&800 &1.332445 &1.333469  & & & &\\
\hline
$S_0 =10$& 50&0.792953 &0.791768  & & & &\\
&100 &0.794123 &0.794123  & & & &\\
&200&0.794515 &0.795156  &0.7961 &0.7958 &0.7960 & 0.7992\\
&400&0.794605  &0.795589  & & & &\\
&800&0.794632  &0.795790  & & & &\\
\hline
$S_0 =11$& 50& 0.445496 &0.444646  & & & &\\
&100&0.446940 &0.446940  & & & &\\
&200& 0.447200 &0.447661  &0.4483 &0.4481 &0.4490 &0.4536\\
&400 &0.447273 &0.447983  & & & &\\
&800 &0.447286 &0.448131  & & & &\\
\hline
$S_0 =12$& 50 &0.241466 &0.241199  & & & &\\
&100 &0.242226 &0.242226  & & & &\\
&200 &0.242360 &0.242546  &0.2428 &0.2427 &0.2430  &0.2502 \\
&400 &0.242393 &0.242678  & & & &\\
&800 & 0.242404 &0.242743 & & & &\\
\hline
\end{tabular}}
\caption{\em \small{Prices of American put options. $S_0=8, 9,
    10,11,12$, $K=10$, $T=0.25$, $r=0.1$, $\delta=0$, $V_0=0.25$,
    $\theta=0.16$, $\kappa=5$, $\rho=0.1$, $\sigma=0.9$.}}
\label{tab4}
\end{table}

\begin{table} [ht] \centering
\footnotesize
{\begin{tabular} {@{}cccccc@{}}\toprule $N_S$ & $S_0 =8$
    & $S_0 =9$ & $S_0 =10$ & $S_0 =11$ & $S_0 =12$   \\
\hline
200 & 2.194914 &2.653543 & 2.280589 &3.181503& 3.204384\\
400 &  2.115892 &2.322611 & 2.380855 &2.237965 &2.423995 \\
800 &2.074822 & 2.140352 &2.165029 &2.178548 &2.044881 \\
\hline
\end{tabular}}
\caption{\em \small{HTFD2-ratio for the price of American put options as the starting point $S_0$ varies. Test parameters are $K=10$, $T=0.25$, $r=0.1$, $\delta=0$, $V_0=0.25$,
    $\theta=0.16$, $\kappa=5$, $\rho=0.1$, $\sigma=0.9$.}}
\label{tab4ratio}
\end{table}

\subsection{European and American barrier options}\label{sect-num-bar}
We study here the continuously monitered barrier options case and we  compare our hybrid tree-finite
difference algorithm  with the numerical results of the method of lines provided in Chiarella et al. \cite{ckm}.
So, we consider European and American up-and-out call options with the following set of parameters: $K=100$, $T=0.5$, $r=0.03$, $\delta=0.05$, $V_0=0.1$, $\theta=0.1$, $\kappa=2$, $\rho=-0.5$.
The up barrier is $H=130$.  We choose  different values for $S_0$:
$S_0=80, 100, 120$.

We also compare with a benchmark value obtained by using
the method of lines, called \textbf{MOL}, with mesh parameters $100, 200, 6400$ (see Chiarella et al. \cite{ckm}).

Table \ref{tab5} and Table \ref{tab6} report European
and American Up-and-Out  option prices respectively, while
Table \ref{tab7} refers to the computational time cost (in seconds) of the various algorithms for the European barrier case.

\begin{table} [ht] \centering
\footnotesize
{\begin{tabular} {@{}ccccc@{}} \toprule & $N_S$ &HTFD1  &HTFD2 & MOL \\
\hline
$S_0=$80& 50 &0.913861 &0.875374 &\\
& 100 &0.893484 &0.893484  &\\
& 200  &0.895127  &0.900893  &0.9029\\
& 400  & 0.897820 &0.902770 &\\
\hline
$S_0=$100& 50 &2.635396 &2.583568 &\\
& 100 &2.606249 &2.606249  &\\
& 200  &2.597363  &2.591857  &2.5908\\
& 400  &2.603679 &2.594134 &\\

\hline
$S_0=$120& 50 &1.417225 &1.438429 &\\
& 100 &1.485704 &1.485704  &\\
& 200  &1.500692  &1.482193  &1.4782\\
& 400  &1.504755 &1.486212 &\\
\hline
\end{tabular}}
\caption{\em \small{Prices of European call up-and-out options. Up
    barrier is $H=130$. $K=100$, $T=0.5$, $r=0.03$, $\delta=0.05$, $V_0=0.1$, $\theta=0.1$, $\kappa=2$, $\rho=-0.5$.}}
\label{tab5}
\end{table}

\begin{table} [ht] \centering
\footnotesize
{\begin{tabular} {@{}ccccc@{}} \toprule & $N_S$ &HTFD1 &HTDF2  \\
\hline
& 50 &0.007& 0.017 \\
& 100  &0.132 &0.132\\
& 200  &0.284 &1.079\\
& 400  &0.535 &8.901\\
\end{tabular}}
\caption{\em \small{Computational times (in seconds) for European barrier
    options.}}
\label{tab7}
\end{table}

\begin{table} [ht] \centering
\footnotesize
{\begin{tabular} {@{}ccccc@{}} \toprule & $N_S$ &HTFD1  &HTFD2 & MOL \\
\hline
$S_0=$80& 50 &1.199802 &1.285959 &\\
& 100 &1.369914 &1.369914  &\\
& 200  &1.400823  &1.396628  &1.4012\\
& 400  &1.400710 &1.401111 &\\
\hline
$S_0=$100& 50 &8.274116 &8.269779 &\\
& 100 &8.286667 &8.286667  &\\
& 200  &8.284054  &8.294226  &8.3003\\
& 400  & 8.283815 &8.296745 &\\
\hline
$S_0=$120& 50 &21.943742 &21.884228 &\\
& 100 &21.820015 &21.820015  &\\
& 200  &21.785274  &21.815989  &21.8216\\
& 400  &21.779648 &21.804518 &\\
\hline
\end{tabular}}
\caption{\em \small{Prices of American call up-and-out options.  Up
    barrier is $H=130$. $K=100$, $T=0.5$, $r=0.03$, $\delta=0.05$, $V_0=0.1$, $\theta=0.1$, $\kappa=2$, $\rho=-0.5$.}}
\label{tab6}
\end{table}

\clearpage

\section{Conclusions}
In this paper, we have introduced a novel numerical method, a hybrid
tree-finite difference method to approximate the Heston model.  The
convergence of the approximating algorithm is studied using a Markov
chain approach.
The numerical results confirm the reliability of the method both in
plain vanilla and barrier options cases. Moreover the numerical
results have shown that the hybrid tree-finite difference method is
very efficient, precise and robust for option pricing in the Heston model.
Finally, our method may be easily generalized to other existing stochastic volatility models, even in the presence of jumps.


\appendix

\section{Appendix}
\subsection{Boundary sensitivity for the implicit finite difference operator}\label{app-boundary}

We study here the behavior of the solution $x=(x_1,\ldots,x_N)^T$ of the two following linear systems
\begin{align}
& A x= \v_1 \label{sys_e1}\\
& A x= \v_N, \label{sys_eM}
\end{align}
where $\v_i$, $i=1,\ldots,N$, denotes the standard orthonormal basis in $\R^N$, i.e.  $(\v_i)_k=0$ for $k\neq i$ and $(\v_i)_i=1$, $i=1,\ldots,N$, and where $A$ has the following general tridiagonal form
\begin{equation}\label{matrixA_gen}
A = \left(
\begin{array}{ccccc}
a_{1} & c_{1} & & & \\
b & a & c & & \\
 & \ddots & \ddots & \ddots &  \\
 &   & b & a & c \\
 &   &        &   b_N & a_N
\end{array}
\right).
\end{equation}
The result we are going to present is due for matrices $A$ as in \eqref{matrixA_gen} and that satisfy the hypotheses (P2)-(P3) in the proof of Proposition \ref{propr_A}, ensuring that they are invertible $M$-matrices (see for instance \cite{mmatrix}).

\begin{proposition}\label{prop_boundary_stima}
Suppose that the matrix $A$ in \eqref{matrixA_gen} satisfies
\begin{equation}\label{A4}
\begin{array}{llllll}
a,a_{1},a_N >0, & b,c\leq 0, & c_{1},b_N\leq 0, &
a>|b|+|c|, & a_{1}>|c_{1}|, & a_{N}>|b_{N}|.
\end{array}
\end{equation}
Assume moreover the following stability conditions on the ``boundary'' values $a_{1}$, $a_{N}$, $c_{1}$ and $b_N$:
\begin{equation}\label{stab_bordo}
\begin{array}{lcl}
\Frac{|bc_1|}{a_1}<z_+ &\mbox{ and }& \Frac{|b_Nc|}{a_N}<z_+,
\end{array}
\end{equation}
where $z_+=(a+\sqrt{a^2-4|bc|})/2$.
Then the solution $x$ of \eqref{sys_e1} is defined by a sequence
$\{x_k\}_{k=1,\ldots,N}$ of positive terms and there exists a positive value $\tilde \gamma^*>|b|$ such that, for $k=2,\ldots,N-1$
\begin{equation}\label{stima_e1}
\begin{array}{lcr}
x_{N+1-k} \leq x_1\, \left(\Frac{|b|}{\tilde \gamma^*}\right)^{N-k} & \mbox{and} & x_N\leq x_1\,\Frac{|b_N|}{a_N}\left(\Frac{|b|}{\tilde \gamma^*}\right)^{N-2}.
\end{array}
\end{equation}
Similarly, for the solution $x$ of \eqref{sys_eM} it holds $x_k>0$, for all $k=1,\ldots,N$ and there exists
a positive value $\gamma^*>|c|$ such that for $k=2,\ldots,N-1$,
\begin{equation}\label{stima_eM}
\begin{array}{lcr}
x_k \leq x_N\,\left(\Frac{|c|}{\gamma^*}\right)^{N-k} & \mbox{and} & x_1\leq x_N\, \Frac{|c_1|}{a_1}\left(\Frac{|c|}{\gamma^*}\right)^{N-2}.
\end{array}
\end{equation}
\end{proposition}
\proof
Let us start by estimating first the solution $x$ of system \eqref{sys_eM}.
By applying the Thomas algorithm,
also known as tridiagonal matrix algorithm \cite{thomas}, the solution of \eqref{sys_eM} is given by back substitutions:
$$
\begin{array}{ccc}
x_N=\Frac{1}{\gamma_N}, &
 x_k = \Frac{|c|}{\gamma_k}x_{k+1}\mbox{ for } k=N-1,\ldots,2, &
x_{1}=\Frac{|c_{1}|}{\gamma_{1}}x_{2},
\end{array}
$$
where the coefficients $\gamma_k$ are recursively defined by
\begin{equation}\label{gammas}
\begin{array}{cccc}
\gamma_{1}=a_{1}, &  \gamma_{2}=a-\Frac{|bc_{1}|}{\gamma_{1}}, &
\gamma_k = a -\Frac{|bc|}{\gamma_{k-1}}\mbox{ for }  k=3,\ldots,N-1, &
\gamma_{N}=a_N - \Frac{|b_{N}c|}{\gamma_{N-1}}.
\end{array}
\end{equation}
It is easy to verify that under assumptions \eqref{A4},
for $k=3,\ldots,N-1$, the sequence $\{\gamma_k\}$ has two strictly positive fixed points $z_{\pm}=(a\pm\sqrt{a^2-4|bc|})/2$. Moreover, $z_-$ is an unstable fixed point while $z_+$ is stable.
By condition \eqref{stab_bordo} and relation
$\gamma_2 = a-|b c_1|/a_1$, we have that
$ \gamma_2>z_-.$
So, starting from $\gamma_2$ the sequence converges to $z_+$ and we have that for $\gamma^*=\min{\{\gamma_{2},z_+\}}$
\begin{equation}\label{cond_gamma_k}
 \gamma_k\geq \gamma_*, \quad k=2,\ldots,N-1.
\end{equation}

By assumptions (A.4), it is easy to verify that $\gamma^*>|c|$ for both cases $\gamma^*=\gamma_2$ and $\gamma^*=z_+$.
%
Moreover, in (A.8) the inequalities $\gamma_{N-1}\geq\gamma^*>|c|$ imply that $\gamma_N>0$.

Going back to the sequence $\{x_k\}_{i=1,\ldots,N}$, we first notice that since $\gamma_N>0$ then $x_N>0$ and 	
accordingly $x_k\geq 0$ for all $k=1,...,N$.
Moreover,
from condition \eqref{cond_gamma_k} we obtain \eqref{stima_eM}. In fact, for $k=2,\ldots,N-1$ we have
$$ x_k = \Frac{|c|}{\gamma_k}x_{k+1}\leq \Frac{|c|}{\gamma^*} x_{k+1} \leq \ldots \leq \left(\Frac{|c|}{\gamma^*}\right)^{N-k}x_N$$
and thus
$$ x_1=\Frac{|c_{1}|}{\gamma_{1}}x_{2}\leq\Frac{|c_{1}|}{a_{1}}\left(\Frac{|c|}{\gamma^*}\right)^{N-2}x_N.$$

To obtain the estimate \eqref{stima_e1} we introduce the $N\times N$ matrix $U$ satisfying $U\v_i=\v_{N+1-i}$, $i=1,\ldots,N$, so
$$
U = \left(
\begin{array}{cccc}
0 &  & 0 & 1 \\
 &  & 1 & \\
 &  \mbox{\rotatebox[origin=c]{90}{$\ddots$}} &  &  \\
1 & 0  &        &   0
\end{array}
\right).
$$
Since $U\v_{N}=\v_{1}$ and $UU=I$ (i.e. $U^{-1}=U$), to compute \eqref{stima_e1} we use that
$$ Ax=\v_1 \Leftrightarrow \tilde A \tilde x = \v_N,$$
where
$$ \tilde A = UAU =
\left(
\begin{array}{ccccc}
a_{N} & b_{N} & & & \\
c & a & b & & \\
 & \ddots & \ddots & \ddots &  \\
 &   & c & a & b \\
 &   &        &   c_{1} & a_{1}
\end{array}
\right)
$$
and
$$ \tilde x = U x = (x_N,x_{N-1},\ldots,x_{1})^T.$$
So, following the same reasoning as above, we get
$\tilde\gamma_2=a-\frac{|b_Nc|}{a_N}$ and
$\tilde\gamma^*=\min{(\tilde\gamma_2,z_+)}>|b|$ such that
$$\tilde x_k \leq \tilde x_N\,\left(\Frac{|b|}{\tilde\gamma^*}\right)^{N-k},\quad k=2,\ldots,N-1$$
i.e.
$$ x_{N+1-k} \leq x_1\,\left(\Frac{|b|}{\tilde\gamma^*}\right)^{N-k}, \quad k=2,\ldots,N-1.$$
\cvd

\begin{remark}\label{bordo-A}
Assume that $A$ has the form \eqref{matrixA_gen} with $a> 0$, $b,c< 0$, $a+b+c=1$ and $a_1=a_N=a$, $c_1=b_N=1-a$ - this is actually the type of matrix to which we apply Proposition \ref{prop_boundary_stima}, see \eqref{matrixA}. One can easily check that both \eqref{A4} and the the boundary requirements in \eqref{stab_bordo} hold, so Proposition \ref{prop_boundary_stima} can be applied. Moreover, estimates \eqref{stima_e1} and \eqref{stima_eM} can be rewritten as follows: for $k=2,\ldots, N-1$,
\begin{eqnarray*}
&|(A^{-1}\v_1)_{N+1-k}|\leq \Big(\displaystyle\frac{|b|}{\tilde\gamma^*}\Big)^{N-k}
&\quad\mbox{with}\quad
\tilde\gamma^{*}=\min\Big(a-\frac{|c(1-a)|}a,\frac{a+\sqrt{a^2-4|bc|}}2\Big),\\
&|(A^{-1}\v_N)_k|\leq \Big(\displaystyle\frac{|c|}{\gamma^*}\Big)^{N-k}
&\quad\mbox{with}\quad
\gamma^{*}=\min\Big(a-\frac{|b(1-a)|}a,\frac{a+\sqrt{a^2-4|bc|}}2\Big).
\end{eqnarray*}
In fact, as for the second inequality, \eqref{stima_eM} gives $|(A^{-1}\v_N)_k|\leq x_N (\frac{|c|}{\gamma^*})^{N-k}$, where $x_N=\frac 1{\gamma_N}$ and $\gamma_N$ is defined in \eqref{gammas}, together with $\gamma_1,\ldots,\gamma_{N-1}$. Since $\gamma_{N-1}>|c|$, \eqref{gammas} gives
$$
\gamma_{N}=a - \Frac{|(1-a)c|}{\gamma_{N-1}}
\geq a-|1-a|.
$$
But $1-a=b+c\leq 0$, so $\gamma_{N}\geq 1$. This implies $x_N\leq 1$ and then $|(A^{-1}\v_N)_k|\leq(\frac{|c|}{\gamma^*})^{N-k}$. Similarly, one has $\tilde\gamma_N\geq 1$, and the first inequality holds as well.
\end{remark}

\begin{remark}\label{bordo-A-dirichlet}
Suppose now that $A$ has the form \eqref{matrixA_gen} with $a> 0$, $b,c< 0$, $a+b+c=1$ and $a_1=a_N=1$, $c_1=b_N=0$ - this is the matrix in \eqref{matrixA-dirichlet}, that is the matrix one has to deal with when the boundary conditions are of a time-independent Dirichlet type. Here, \eqref{A4} and \eqref{stab_bordo} both hold, so Proposition \ref{prop_boundary_stima} can be applied.
Moreover, one immediately gets $\gamma_N=\tilde\gamma_N=1$, $\tilde \gamma_2=\gamma_2=a<z_+$, so that $\min(\tilde\gamma_2,z_+)=\min(\gamma_2,z_+)=a$. Therefore, for $k=2,\ldots, N-1$,
$$
|(A^{-1}\v_1)_{N+1-k}|\leq \Big(\displaystyle\frac{|b|}{a}\Big)^{N-k}
\quad\mbox{and}\quad
|(A^{-1}\v_N)_k|\leq \Big(\displaystyle\frac{|c|}{a}\Big)^{N-k}.
$$
\end{remark}

%

\subsection{The finite difference operators on the infinite grid}\label{app-infinity}

For fixed $\Dy>0$ and $y_0\in\R$, we consider the grid $\mathcal{Y}=\{y_k\}_{k\in\Z}$ on $\R$ by setting
$$
y_k=y_0+k\Dy,\quad k\in\Z.
$$
Any real function $\varphi$ on $\mathcal{Y}$ can be seen as a sequence on $\R$. So, as usual we set $\R^\Z=\{x=(x_k)_{k\in\Z}\,:\, x_k\in\R\ \forall k\}$,and we think of $x\in\R^\Z$ as ``column-type'' vectors. We also  set $\varphi(y)$, $y\in\mathcal{Y}$, as the point in $\R^\Z$ defined as
$$
(\varphi(y))_k=\varphi(y_k),\quad k\in\Z.
$$
We split our discussion in two parts: the case $v>\epsilon$, giving the implicit finite difference operator, and $v\leq \epsilon$, related to the explicit finite difference operator

\subsubsection{The case $v>\epsilon$}

For the implicit in time approximation, equation \eqref{discr_eq} can be written as $u^n=Au^{n+1}$, in which $u^n=(u^n_j)_{j\in\Z}$ is the unknown, $u^{n+1}=(u^{n+1}_j)_{j\in\Z}$ is given and $A=(A_{kj})_{k,j\in\Z}$ is the infinite dimensional matrix  given by
$$
A_{kj}=\left\{
\begin{array}{ll}
-\beta+\alpha & \mbox{if } j=k-1\\
1+2\beta & \mbox{if } j=k\\
-\beta-\alpha & \mbox{if } j=k+1
\end{array}
\right.
\quad\mbox{and $A_{kj}=0$ for $|j-k|> 1$.}
$$
$A$ is actually a linear operator on $\R^\Z$:  by using the standard row/column product, we get $A\,:\,\R^{\Z}\to\R^\Z$  and  $y=Ax$  if and only if $y_k=(Ax)_k=\sum_{j\in\Z}A_{kj}x_j
=(-\beta+\alpha)x_{k-1}+(1+2\beta)x_k+(-\beta-\alpha)x_{k+1}$, $x\in\R^\Z$. So, we write
$$
A=(1+2\beta)I-T
$$
where $I$ denotes the identity map over $\R^\Z$ and $T\,:\,\R^{\Z}\to\R^\Z$ is the linear operator defined through the infinite dimensional matrix
$$
T_{kj}=\left\{
\begin{array}{ll}
\beta-\alpha & \mbox{if } j=k-1\\
\beta+\alpha & \mbox{if } j=k+1
\end{array}
\right.
\quad\mbox{and $T_{kj}=0$ for $|j-k|\neq 1$.}
$$
Throughout this section, we assume
$$
\beta>|\alpha|\geq 0.
$$
This gives that $T\geq 0$, in the sense that $T_{ij}\geq 0$ for every $i,j\in\Z$.

First, we want to show that $A$ is an invertible operator, at least over a nice subset of $\R^\Z$. Secondly, we want to show that the inverse is a stochastic infinite dimensional matrix, that is it defines a Markov transition function over $\Z$.

So, for $a>0$, we set
$$
\la=\Big\{x\in\R^\Z\,:\,\sum_{k\in\Z}a^{2|k|}|x_k|^2<\infty\Big\}.
$$
\begin{remark}\label{app-boh}
If $a<1$ then $\la$ contains all the sequences with polynomial growth: for any fixed $l\in\N$, the sequence $x=(x_k)_k$ such that $|x_k|\leq C(1+|k|^l)$, $k\in\Z$, belongs to $\la$. Moreover, $\ell^2_{a_1}\subset \ell^2_{a_2}$ if $a_1\leq a_2\leq 1$.
Finally, we notice that, for $a<1$,  $\la$ contains also special sequences with exponential growth, namely points $x\in\R^\Z$ such that $|x_k|\leq C(1+e^{\gamma |k|})$ for every $\gamma<-2\log a$.

\end{remark}

On $\la$, we define the inner product and the norm by
$$
\<x,y\>_a=\sum_{k\in\Z}a^{2|k|}x_ky_k
\quad\mbox{and}\quad
|x|^2_a=\sum_{k\in\Z}a^{2|k|}x_k^2
$$
respectively. When $a=1$ one has
$$
\ell^2_1=\Big\{x\in\R^\Z\,:\,\sum_{k\in\Z}|x_k|^2<\infty\Big\},
$$
which is the standard $\ell^2$ space of sequences.

For $a>0$ the map
\begin{equation}\label{app-Da}
D_a\,:\,\la \to\ell^2,\quad (D_ax)_k=a^{|k|}x_k, \quad k\in\Z,
\end{equation}
is well defined and invertible, with $D_a^{-1}=D_{a^{-1}}$. Moreover, since
\begin{equation}\label{app-Da1}
\<x,y\>_a=\<D_ax,D_ay\>,\quad x,y\in\la,
\end{equation}
$\<\cdot,\cdot\>$ denoting the standard scalar product in $\ell^2$, $D_a$ is an isometry between $\la$ and  $\ell^2$. Therefore, the Hilbert structure of $\ell^2$ can be transferred  to $\la$ through $D_a$, so that the space $\la$, endowed with the scalar product $\<\cdot,\cdot\>_a$, is a Hilbert space.

Since the objects of our interest will belong to $\la$ with $a<1$, we study the invertibility of $A$ when it is restricted on $\la$.
\begin{proposition}
Assume that
$$
1>a>\frac{2\beta}{1+2\beta}. 
$$
Then $A\,:\,\la\to \la$.
Moreover, $A$ is invertible and the inverse
$A^{-1}$ can be identified with an infinite dimensional matrix $(\Pi_{i,j})_{i,j\in\Z}$ that does not depend on the choice of $a$: for every $x\in\la$,
$$
(A^{-1}x)_k=\sum_{j\in\Z}\Pi_{kj}x_j\quad\mbox{for every } k\in\Z.
$$
Finally, $\Pi$ defines a Markov transition function, that is  $\Pi_{kj}\geq 0$ and $\sum_{j\in\Z}\Pi_{kj}=1$ for every $k,j\in\Z$.
\end{proposition}

\textbf{Proof.} For $x\in\la$, we set $(I_{\pm}x)_k=x_{k\pm 1}$, $k\in\Z$. So, we have
$$
Tx=(\beta-\alpha)I_-x+(\beta+\alpha)I_+x.
$$
Straightforward computations give $|I_-x|^2_a
=\frac 1{a^2}\sum_{k\leq 0}a^{2|k|}x_k^2
+a^2\sum_{k\geq 1}a^{2|k|}x_k^2$, so that
$|I_-x|_a\leq \frac 1{a}|x|_a$. And similarly, one gets
$|I_+x|_a\leq \frac 1{a}|x|_a$.
Therefore, 
$$
|Tx|_a\leq (\beta-\alpha)|I_-x|_a+(\beta+\alpha)|I_+x|_a
\leq \frac {2\beta}{a}\,|x|_a
$$
and this gives $T\,:\,\la\to\la$, so that $A\,:\,\la\to\la$. Moreover, one has $\|T\|_a\leq \frac {2\beta}{a}$, $\|\cdot\|_a$ denoting the standard operator norm on $\la$. As a consequence one gets
$$
\|T^n\|_a\leq \Big(\frac {2\beta}{a}\Big)^n\quad\mbox{for every }n\geq 0.
$$
We prove now that $A$ is invertible on $\la$. For $N\in\N$, we set $S_N\,:\,\la\to\la$ through
$$
S_N=\frac 1{1+2\beta}\sum_{n=0}^N\frac 1{(1+2\beta)^n}\,T^n.
$$
We notice that
$$
\|S_N\|_a\leq \frac 1{1+2\beta}\sum_{n=0}^N\frac 1{(1+2\beta)^n}\,\|T^n\|_a
\leq \frac 1{1+2\beta}\sum_{n=0}^N\Big(\frac {2\beta}{a(1+2\beta)}\Big)^n.
$$
Since $\frac {2\beta}{a(1+2\beta)}<1$, the series converges and the limit (linear) operator $S\,:\,\la\to\la$ exists. Moreover, we can write
$$
S=\frac 1{1+2\beta}\sum_{n=0}^\infty\frac 1{(1+2\beta)^n}\,T^n.
$$
Let us prove that $S=A^{-1}$. For every $N$, one has
$$
S_NA
=S_N\big((1+2\beta)I-T\big)
=I-\frac 1{(1+2\beta)^{n+1}}\,T^{n+1}
$$
and by taking the limit as $N\to\infty$ one gets $SA=I$. Similarly, one proves that $AS=I$, so that $S=A^{-1}$ on $\la$.

Now, let $\{\e_i\}_{i\in\Z}\subset\ell^2$ denote the standard orthonormal basis: $(\e_i)_k=0$ for $k\neq i$ and $(\e_i)_i=1$. Since $\ell^2=\la$ with $a=1>\frac{2\beta}{1+2\beta}$, we can set
$$
\Pi_{kj}=\<\e_k,S\e_j\>,\quad k,j\in\Z.
$$
We prove now  that $\Pi$ is actually the matrix we are looking for: if $x\in\la$ then $(Sx)_k=\sum_{j\in\Z}\Pi_{kj}x_j$, $k\in\Z$.

Let $D_a$ denote the linear operator in \eqref{app-Da}. Since $D_a^{-1}=D_{a^{-1}}$,  \eqref{app-Da1} immediately gives that $\{D_{a^{-1}}\e_i\}_i$ is an orthonormal basis in $\la$. Moreover, for $x\in\la$ one has
$$
x=\sum_{j\in\Z}x_j\e_j
=\sum_{j\in\Z}a^{|j|}x_ja^{-|j|}\e_j
=\sum_{j\in\Z}(D_ax)_jD_{a^{-1}}\e_j
$$
and since $Sx\in\la$, we can also write
$$
\<D_{a^{-1}}\e_k,Sx\>_a=a^{|k|}(Sx)_k.
$$
So,
\begin{align*}
(Sx)_k
&=a^{-|k|}\<D_{a^{-1}}\e_k,Sx\>_a
=a^{-|k|}\sum_{j\in\Z}(D_ax)_j\<D_{a^{-1}}\e_k,SD_{a^{-1}}\e_j\>_a\\
&=a^{-|k|}\sum_{j\in\Z}a^{|j|}x_j\<D_{a^{-1}}\e_k,SD_{a^{-1}}\e_j\>_a
=\sum_{j\in\Z}a^{|j|-|k|}\<D_{a^{-1}}\e_k,SD_{a^{-1}}\e_j\>_ax_j.
\end{align*}
Therefore, we get  $(Sx)_k=\sum_{j}\Pi_{kj}x_j$ with
$$
\Pi_{kj}
=a^{|j|-|k|}\<D_{a^{-1}}\e_k,SD_{a^{-1}}\e_j\>_a.
$$
We show now that actually $\Pi$ is independent of $a$ and moreover $\Pi_{kj}=\<\e_k,S\e_j\>$. Since $D_{a^{-1}}\e_j=a^{-|j|}\e_j$ and $\e_j\in\la$, using \eqref{app-Da1} and the fact that $D_ax\in\ell^2$ if $x\in\la$, we can write
\begin{align*}
\Pi_{kj}
&=a^{-|k|}\<D_{a^{-1}}\e_k,S\e_j\>_a
=a^{-|k|}\<D_{a^{-1}}\e_k,D_{a^{-1}}D_{a}S\e_j\>_a
=a^{-|k|}\<\e_k,D_{a}S\e_j\>\\
&=a^{-|k|}(D_{a}S\e_j)_k
=a^{-|k|}a^{|k|}(S\e_j)_k
=\<e_k,S\e_j\>.
\end{align*}
In order to show that $\Pi$ is a Markov transition function on $\Z$, we proceed as follows.
One has
$$
\Pi_{kj}=\<\e_k,S\e_j\>
=\lim_{N\to\infty}\<\e_k,S_N\e_j\>
=\lim_{N\to\infty}\frac 1{1+2\beta}\sum_{n=0}^N\frac 1{(1+2\beta)^n}(T^n)_{i,j}\geq 0
$$
because the entries of $T$ are all $\geq 0$, and so it is for $T^n$, for every $n$.
Finally, let $\Vunita\in\R^\Z$ be the unit vector, that is  $(\Vunita)_k=1$ for every $k\in\Z$. Notice that $\Vunita\in\la$ for every $a<1$ and $T\Vunita=2\beta\Vunita$. Then
$S\Vunita$ is well defined and
$$
S\Vunita=\frac 1{1+2\beta}\sum_{n=0}^\infty\Big(\frac{2\beta}{1+2\beta}\Big)^n\Vunita
=\Vunita
$$
and this automatically gives $\sum_{j\in\Z}\Pi_{kj}=1$ for every $k\in\Z$.
\cvd

For each $l\in \N$, we consider the polynomial $(y-y_i)^l$ and we call $\psi^i_l(y)\in\R^\Z$ the associated (vector) function of  $y\in\mathcal{Y}$:
\begin{equation}\label{app-psi}
\big(\psi^i_l(y)\big)_k =
(y_k-y_i)^l=\Dy^l(k-i)^l, \quad k\in\Z.
\end{equation}
By Remark \ref{app-boh}, one has $\psi^i_l(y)\in\la$ for every fixed $a<1$, $i\in\Z$ and $l\in\N$, so the quantity $A^{-1}\psi^i_l(y)$ makes sense. We also notice that $\psi^i_0(y)=\Vunita$, so that $A^{-1}\psi^i_0(y)=A^{-1}\Vunita=\Vunita$.

As already developed in Section \ref{sect-implicit}, we need to deal with $A^{-1}\psi^i_l(y)$ for $l\leq 4$ and $i\in\Z$.

First of all we have

\begin{lemma}\label{app-Apsi}
Let $\psi^i_l(y)$ be defined in \eqref{app-psi}. Then for every $l\in\N$ and $i\in\Z$ one has
$$
A\psi^i_l(y)=\psi^i_l(y)-\sum_{j=0}^{l-1}\coeffbin{l}{j}a_{l-j}\Dy^{l-j}\psi^i_j(y).
$$
where
\begin{equation}\label{app-an}
a_n=(\beta-\alpha)(-1)^n+(\beta+\alpha),\quad n\in\N,
\end{equation}
that is $a_n=2\beta$ if $n$ is even and $a_n=2\alpha$ if $n$ is odd.
\end{lemma}

The proof is identical to the one of Proposition \ref{Apsi}, so we omit it. We only notice that, due to the use of the grid on the whole real line, we do not have any contribution from the boundary - see the terms $b_{l,i}^{\pm M}\e_{\pm M}$ in \eqref{Apsi_form}. This gives that the inverse of $A$ on any polynomial $\psi^i_l(y)$ is much simpler, and in fact we have

\begin{proposition}\label{app-prop-psi}
For $l\geq 1$ let $\gamma_{l, k}$, $k=0,1,\ldots,l$, be iteratively (backwardly) defined as follows:
$$
\gamma_{l,k}=
\coeffbin{l}{k}a_{l-k}\Dy^{l-k}+
\sum_{j=k+1}^{l-1}\gamma_{l,j}\coeffbin{j}{k}a_{j-k}
\Dy^{j-k},\quad
k=l-1,\ldots,0,
$$
where the coefficients $a_n$ are given in \eqref{app-an}. Then
$$
A^{-1}\psi^i_l(y)=\psi^i_l(y)+\sum_{j=0}^{l-1}\gamma_{l,j}\psi^i_j(y).
$$
\end{proposition}

For the proof, we again refer to the finite case, that is Proposition \ref{prop-psi}.

As a consequence, for $l=1,2,4$ and $i\in\Z$, the formulas for $(A^{-1}\psi^i_l(y))_i$ are identical to the ones for the finite grid (see \eqref{pippoA}) except for the boundary terms, which are null here. In fact, by inserting the formula \eqref{alphabeta} for $\beta$ and $\alpha$, we get
\begin{equation}\label{app-pippoA}
\begin{array}{rl}
(A^{-1}\psi_1^i(y))_i
=&h\mu_Y(v),\\
(A^{-1}\psi_2^i(y))_i
=&h\bar\rho^2v+2h\Dy\mu_Y(v),\\
(A^{-1}\psi_4^i(y))_i
=&
h\Dy^2\bar\rho^2v+8h^2\Dy^2\mu_Y(v)^2+24h^3\mu_Y(v)^3
+6h^2\bar\rho^4v^2+24\frac{h^4}{\Dy}\,\bar\rho^2v\mu_Y(v)^3,
\end{array}
\end{equation}
for every $i\in\Z$.
\subsubsection{The case $v\leq \epsilon$}

The operator associated to the the explicit in time approximation is given by
$$
C_{kj}=\left\{
\begin{array}{ll}
\beta+2|\alpha|\I_{\{\alpha<0\}} & \mbox{if } j=k-1\\
1-2\beta-2|\alpha| & \mbox{if } j=k\\
\beta+2|\alpha|\I_{\{\alpha>0\}} & \mbox{if } j=k+1
\end{array}
\right.
\quad\mbox{and $C_{kj}=0$ for $|j-k|> 1$.}
$$
$\alpha$ and $\beta$ being given in \eqref{alphabeta}, and the solution $u^n\in\R^\Z$ is given by $u^n=Cu^{n+1}$. The infinite dimensional matrix $C$ defines a Markov transition function if and only if
$$
  2\beta+2|\alpha|\leq 1.
$$
Here, nothing changes with respect to what developed in Section \ref{sect-explicit}. So, for every $i\in\Z$ we get the same formulas in \eqref{pippoC}, that is
\begin{equation}\label{app-pippoC}
\begin{array}{rl}
(C\psi_1^i(y))_i
&=h\mu_Y(v),\\
(C\psi_2^i(y))_i
&=h\rho^2 v+ h\Dy|\mu_Y(v)|,\\
(C\psi_4^i(y))_i
&=h\Dy^2\rho^2 v + h\Dy^3|\mu_Y(v)|,
\end{array}
\end{equation}
in which we have inserted the formulas for $\alpha$ and $\beta$ in \eqref{alphabeta}.

\begin{remark}
By using \eqref{app-pippoA} and \eqref{app-pippoC}, we can prove the convergence result in Theorem \ref{th-conv} also in the case of the infinite grid. The proof is identical, even simpler because we do not have here contributions from the boundary.
\end{remark}


\begin{thebibliography}{}


\bibitem{al} \textsc{A. Alfonsi} (2010): High order discretization schemes for the CIR
    process: application to affine term structure and Heston models,  {\it
      Mathematics of Computation}, \textbf{79}, 209-237.

\bibitem{ak} \textsc{K. Amin, A. Khanna} (1994): Convergence of American option values from discrete-to continuous-time financial models , {\it Mathematical Finance}, {\bf 4}, 289-304.

\bibitem{bib:acz} \textsc{E. Appolloni, L. Caramellino, A. Zanette} (2014): A robust tree method
for pricing American options with CIR stochastic interest rate. {\it IMA Journal of Management Mathematics}, to appear, doi:10.1093/imaman/dpt030; \texttt{ArXiv:1305.0479}.

\bibitem{mmatrix} \textsc{A. Berman, R. J. Plemmons} (1994): Nonnegative matrices in the mathematical sciences, {\it Society for Industrial and Applied Mathematics (SIAM)}, Philadelphia, PA.

\bibitem{bill}
\textsc{P. Billingsley} (1968):  \textit{Convergence  of  probability  measures}.  Wiley,  New  York.


\bibitem{bs} \textsc{M.J. Brennan, E.S. Schwartz} (1976): The pricing of
    equity-linked life insurance policies with an asset value guarantee, {\it
      Journal of Financial Economics}, {\bf 3}, 195-213.

\bibitem{ckm} \textsc{C. Chiarella, B.Kang, G.H. Meyer} (2012): The evaluation
  of barrier option prices under stochastic volatility, {\it
      Computers and Mathematics with Applications}, {\bf 64}, 2034-2048.

\bibitem{ClarkeParrott}
\textsc{N. Clarke, K. Parrott.} (1999):
Multigrid for {A}merican option pricing with stochastic volatility.
{\em Applied Mathematical Finance}, \textbf{6}, 177--195.


\bibitem{cir} \textsc{J. C. Cox, J. Ingersoll, S. Ross} (1985): A theory of the term structure of interest rates, {\it Econometrica}, {\bf 53}, 385-407.

\bibitem{dfl} \textsc{V.D'Halluin, P.A.Forsyth, G.Labahn} (2005): A semi-Lagrangian
    Approach for American Asian options under jump-diffusion, {\em
      Siam J.Sci.Comp.} {\bf 27}, 315-345.

\bibitem{ek} \textsc{S.N. Ethier, T. Kurtz} (1986): \textit{Markov processes: characterization and convergence}. John Wiley \& Sons, New York.

\bibitem{fo} \textsc{F. Fang, C.W. Oosterlee} (2011): A Fourier-based
  valuation method for Bermudan and barrier options under Heston's
  model, \textit{SIAM J. Fin. Math}, \textbf{2}, 439--463.

\bibitem{FeehanPop1}
\textsc{P.M.N. Feehan, C.A. Pop} (2013):
On the martingale problem for degenerate-parabolic partial differential operators with unbounded coefficients and mimicking theorem for It\^o processes. \emph{Transactions of The American Mathematical Society}, to appear; \texttt{ArXiv 1211.46362}.

\bibitem{FeehanPop2}
\textsc{P.M.N. Feehan, C.A. Pop} (2013):
A Schauder approach to degenerate parabolic partial differential equations with unbounded coefficients. Preprint \texttt{ArXiv 1211.46362}.

\bibitem{FinucaneTomas}
\textsc{T.J. Finuance, M.J. Tomas} (1996):
American stochastic volatility call option pricing: a lattice based approach.
{\em Review of Derivatives Research}, \textbf{1}, 183--201.

\bibitem{FlorescuViens1}
\textsc{I. Florescu, F. Viens} (2005):
A binomial tree approach to stochastic volatility driven model of the
stock price. {\em An. Univ. Craiova Ser. Mat. Inform.}, \textbf{32}, 126--142.


\bibitem{FlorescuViens2}
\textsc{I. Florescu, F. Viens} (2008):
Stochastic volatility: option pricing using a multinomial recombining tree.
{\em Applied Mathematical Finance Journal}, \textbf{15}, 151--181.

\bibitem{GuanXiaoqiang}
\textsc{L.K. Guan, G. Xiaoqiang} (2000):
Pricing {A}merican options with stochastic volatility: {E}vidence
  from {S}\&{P} futures options.
{\em Journal of Futures Markets}, \textbf{20}, 625--659.

\bibitem{hhv} \textsc{T. Haentjens, K.J. in't Hout, K. Volders} (2010): ADI schemes with
Ikonen-Toivanen splitting for pricing American put options in the
Heston model. In: \textit{Numerical Analysis and Applied Mathematics,
eds. T. E. Simos et. al., AIP Conf. Proc.} 1281, 231-234.

\bibitem{hes} \textsc{S.L. Heston} (1993): A Closed-Form Solution for Options with Stochastic Volatility with Applications to
Bond and Currency Options, \textit{Review of Financial Studies}, \textbf{6}, 327-343.

\bibitem{HilliardSchwartz}
\textsc{J.E. Hilliard, A. Schwartz} (1996):
Binomial option pricing under stochastic volatility and correlated
  state variables.
{\em Journal of Derivatives}, 23--39.

\bibitem{HullWhite}
\textsc{J. Hull, A. White} (1987): The pricing of options on assets with stochastic volatilities. {\em The Journal of Finance}, \textbf{42}, 281--300.

\bibitem{it} \textsc{S. Ikonen, J. Toivanen} (2009): Operator splitting methods for pricing American options under stochastic volatility, \textit{Numer. Math.},
\textbf{113},  299-324.

\bibitem{hf}  \textsc{K.J. in't Hout, S. Foulon} (2010): ADI finite difference schemes for
option pricing in the Heston model with
correlation.\textit{Int. J. Numer. Anal. Mod. }, \textbf{7}, 303-320.

\bibitem{kd} \textsc{H. Kushner, P.G. Dupuis} (1992): \textit{Numerical methods for stochastic
  control problems in continous time.} Springer Verlag.


\bibitem{lp} \textsc{D. Lamberton, G. Pag\`es} (1990): Sur l'approximation des r\'eduites. \textit{Ann. Inst. H. Poincar\'e, Probab. et Statistiques} \textbf{26}, 331-355.

\bibitem{LeisenTreeHeston}
\textsc{D.P.J. Leisen} (2000):
Stock evolution under stochastic volatility: A discrete approach.
{\em Journal of Derivatives}, \textbf{8}, 8--27.

\bibitem{ls}
\textsc{F.A. Longstaff, E.S. Schwartz} (2001): Valuing American options
by simulations: a simple least squares approach.  \it The Review of
Financial Studies, \bf 14\rm, 113-148.

\bibitem{nr} \textsc{D.B. Nelson, K. Ramaswamy} (1990): Simple binomial processes as diffusion approximations in financial models. \textit{The Review of Financial Studies},  \textbf{3}, 393-430.

\bibitem{vn} \textsc{H. Nieuwenhuis, M. Vellekoop} (2009): A tree-based method to price American Options in the
  Heston Model. \textit{The Journal of Computational Finance},  \textbf{13}, 1--21.

\bibitem{Oosterlee}
\textsc{C.W. Oosterlee} (2003):
On multigrid for linear complimentarity problems with application to
  {A}merican-style options.
{\em Electronic Transactions on Numerical Analysis}, \textbf{15}, 165--185.

\bibitem{pr} \textsc{PREMIA}: An Option Pricer. {\tt http://www.premia.fr}

\bibitem{SteinStein}
\textsc{E. Stein, J. Stein} (1991): Stock price distributions with stochastic volatility: an analytic approach. {\em Review of Financial Studies}, \textbf{4}, 727--752

\bibitem{strikwerda} \textsc{J. C. Strikwerda} (2004): Finite difference schemes and partial differential equations,  {\it Society for Industrial and Applied Mathematics (SIAM)}, Philadelphia, PA.

\bibitem{sw}
\textsc{D.W. Stroock,  S.R.S.  Varadhan} (1979): \textit{Multidimensional  Diffusion  Processes}. Springer, Berlin.

\bibitem{thomas}
\textsc{L. H. Thomas} (1949): Elliptic Problems in Linear Differential Equations over a Network,  {\it Watson Sci. Comput. Lab Report}, Columbia University, New York.

\bibitem{tian1} \textsc{Y. Tian} (1992): A simplified binomial approach to the pricing
    of interest-rate contingent claims. {\it Journal of Financial Engineering},
    {\bf 1}, 4-37.

\bibitem{tian2} \textsc{Y. Tian} (1994): A reexamination of lattice procedures for interest rate-contingent claims. {\it Advances in Futures and Options Research}, {\bf 7}, 87-110.

\bibitem{ForsythVetzalZvan}
\textsc{R. Zvan, P. Forsyth, K. Vetzal} (1998):
A penalty method for {A}merican options with stochastic volatility.
{\em J. Comp. Appl. Math.}, \textbf{92}, 199--218.

\end{thebibliography}
\end{document}